# Assessing Rate limits Using Behavioral and Neural Responses of Interaural-Time-Difference Cues in Fine-Structure and Envelope

**Hongmei Hu**[1,2,3*]**, Stephan Ewert**[2,3]**, Birger Kollmeier**[2,3]**, Deborah Vickers**[1]

[1]SOUND Lab, Cambridge Hearing Group, Department of Clinical Neuroscience, Cambridge University, Cambridge, UK

[2]Department of Medical Physics and Acoustics, University of Oldenburg, Oldenburg, Germany

[3]Cluster of Excellence "Hearing4all", University of Oldenburg, Oldenburg, Germany

**\*Correspondence:**

Corresponding Author

hh594@cam.ac.uk, hongmei.hu@uol.de

## Abstract

The objective was to determine the effect of pulse rate on the sensitivity to use interaural-time-difference (ITD) cues and to explore the mechanisms behind rate-dependent degradation in ITD perception in bilateral cochlear implant (CI) listeners using CI simulations and electroencephalogram (EEG) measures. To eliminate the impact of CI stimulation artifacts and to develop protocols for the ongoing bilateral CI studies, upper-frequency limits for both behavior and EEG responses were obtained from normal hearing (NH) listeners using sinusoidal-amplitude-modulated (SAM) tones and filtered clicks with changes in either fine structure ITD or envelope ITD. Multiple EEG responses were recorded, including the subcortical auditory steady-state responses (ASSRs) and cortical auditory evoked potentials (CAEPs) elicited by stimuli onset, offset, and changes. Results indicated that acoustic change complex (ACC) responses elicited by envelope ITD changes were significantly smaller or absent compared to those elicited by fine structure ITD changes. The ACC morphologies evoked by fine structure ITD changes were similar to onset and offset CAEPs, although smaller than onset CAEPs, with the longest peak latencies for ACC responses and shortest for offset CAEPs. The study found that high-frequency stimuli clearly elicited subcortical ASSRs, but smaller than those evoked by lower carrier frequency SAM tones. The 40-Hz ASSRs decreased with increasing carrier frequencies. Filtered clicks elicited larger ASSRs compared to high-frequency SAM tones, with the order being 40-Hz-ASSR>160-Hz-ASSR>80-Hz-ASSR>320-Hz-ASSR for both stimulus types. Wavelet analysis revealed a clear interaction between detectable transient CAEPs and 40-Hz-ASSRs in the time-frequency domain for SAM tones with a low carrier frequency.

**Introduction**

Bilateral cochlear implant (CI) users have demonstrated sensitivity to both interaural time difference (ITD) (Egger et al., 2016; Francart et al., 2015; Ihlefeld et al., 2014; Kan et al., 2015) and the interaural level difference (ILD) cues (Best et al., 2011; Stakhovskaya and Goupell, 2017) when stimulated with single or multiple electrode pairs. Their accuracy for localizing sounds in the azimuthal plane (e.g., Litovsky et al., 2009; Nopp et al., 2004) is also improved by exploiting the ILD, and potentially the envelope ITD information (Laback et al., 2004; Seeber and Fastl, 2008). However, there are still substantial performance gaps between bilateral CI users and normal hearing (NH) listeners in various binaural tasks (Grantham et al., 2007; Hu et al., 2018; Litovsky et al., 2012). For example, the average ITD detection thresholds for CI listeners are about 5-10 times higher than thresholds for NH listeners even at low pulse rates. Furthermore, ITD discrimination becomes poor or impossible for pulse rates above 300 pulse per second (pps) for most bilateral CI listeners (Ihlefeld et al., 2015; Kan and Litovsky, 2015; Laback et al., 2015; Laback et al., 2007; van Hoesel, 2007). Such a rate is much lower compared to the upper-frequency limit for pure tones in NH listeners, which is between 1100-1500 Hz (Brughera et al., 2013; Füllgrabe and Moore, 2017, 2018; Grose and Mamo, 2010; Hopkins and Moore, 2010, 2011; Papesh et al., 2017; Ross et al., 2007a; Ross et al., 2007b; Zwislocki and Feldman, 1956). It is also lower than typical pulse rates used in sound processing strategies for delivering sound information in CI processors, which could be 900-5000 pps. To understand the underlying mechanisms of this rate-dependent degradation of ITD sensitivity in bilateral CI users, in addition to the above-mentioned psychoacoustic experiments, various studies have been performed using animal models (Chung et al., 2016; Hancock et al., 2010; Rosskothen-Kuhl et al., 2021; Smith and Delgutte, 2008; Vollmer, 2018) and computer models (e.g., Chung et al., 2015; Colburn et al., 2009; Hu et al., 2023; Hu et al., 2022; Müller et al., 2022) of bilateral CI hearing. However, the dependence of neural ITD sensitivity on pulse rate has not been systematically quantified using non-invasive electroencephalogram (EEG) measures in CI listeners. One of the possible reasons is the process of removing the known CI stimulation artifacts which could distort the responses (e.g., Hofmann and Wouters, 2010; Hu and Dietz, 2015; Hu and Ewert, 2021; Hu et al., 2015). The current study has two research objectives: 1) to develop an EEG test paradigm to predict the ITD sensitivity and the rate limitation in bilateral CI users; 2) to further develop the EEG test procedure as a potential clinical tool for testing binaural hearing, such as for binaural sensitivity screening and bilateral CI fitting. For this latter objective of developing the clinical application there are two key sub-questions: (i) can reliable physiological responses be recorded using a clinically-applicable EEG setup within a limited time period similar to clinical appointment times (typically shorter than 1 hour); (ii) Can multiple brain responses that provide information for different levels in the auditory pathway be measured within one session using multi-information recording techniques? In this preliminary work, EEG responses were recorded from NH participants to exclude the possible effect of CI stimulation artifacts and to develop the protocols for our ongoing bilateral CI studies.

Various binaural EEG and magnetoencephalography (MEG) paradigms have been developed to assess binaural hearing function at different levels of the auditory pathways (e.g., Dobie and Berlin, 1979; Eddins and Eddins, 2018; Gnanateja and Maruthy, 2019; Grose and Mamo, 2012; Hu and Dietz, 2015; Koerner et al., 2020; McAlpine et al., 2016; Ozmeral et al., 2016; Papesh et al., 2017; Ross, 2008; Ross et al., 2007a; Ross et al., 2007b; Shinn-Cunningham et al., 2017; So and Smith, 2020; Ungan et al., 2020; Vercammen et al., 2018). This study measured the ability to detect a change in the interaural phase difference (IPD) or ITD cues. The paradigm was based on the research by Ross et al. (2007a; 2007b), which has been used in psychoacoustic and physiological studies to diagnosis binaural temporal processing abilities, including age-related changes (Füllgrabe et al., 2017; Grose and Mamo, 2010; see review, Moore, 2021; Papesh et al., 2017; Ross, 2018; e.g., Ross et al., 2007a; Ross et al., 2007b). In addition, within the same EEG paradigm, multi-information was recorded and analyzed, including cortical auditory evoked potentials (CAEPs) elicited by stimuli onset/offset, acoustic change complex (ACC) (Martin and Boothroyd, 1999; Ostroff et al., 1998), and subcortical auditory steady-state responses (ASSRs).

Previous behavior studies have reported similar rate limitations for envelope ITD sensitivity of NH listeners (Bernstein and Trahiotis, 2002; Goupell et al., 2009; Henning, 1974; Majdak and Laback, 2009; Monaghan et al., 2015; Nuetzel and Hafter, 1981) and pulse-based ITD sensitivity of bilateral CI users (Ihlefeld et al., 2015; Laback et al., 2007; Poon et al., 2009; van Hoesel, 2007). For setting a benchmark for the upcoming bilateral CI experiments, the upper-frequency limits of both fine-structure ITD ($ITD_{FS}$) and envelope ITD ($ITD_{ENV}$) sensitivities were obtained from the same NH participants.

Three different experiments were performed. In experiment 1, sinusoidal amplitude modulated (SAM) tones with a change in fine structure ITD were used to demonstrate the decreased ability to detect a given ITD above a certain carrier frequency (EEG, Papesh et al., 2017; e.g., MEG, Ross et al., 2007b). The upper limit was suggested to be a potential indication of the binaural hearing deficit since it decreased in both middle-aged and older adults (Ross et al., 2007a). Experiment 1 was designed to validate the test setup, estimate the measurement time, and reproduce the upper carrier frequency of the fine structure ITD sensitivity reported in the literature. Thus the behavioral upper limit of fine structure ITD sensitivity and the physiological responses at different carrier frequencies were measured in the same participants.

Experiment 2 was designed to find out the possible causes for an absent ACC response when evoked purely by a change in envelope ITD. Recently, Ross (2018) used this paradigm to test the envelope ITD change responses using 40 Hz amplitude-modulated sinusoidal (SAM) tones at carrier frequencies of 250, 500, 1000, 2000, and 4000 Hz. For a carrier frequency of 4000 Hz, only two out of 14 participants showed significant responses to the envelope ITD changes. This result is striking because the ACC has raised strong interest in recent years as a cortical objective tool for assessing monaural auditory discrimination capabilities (Calcus et al., 2022; Han and Dimitrijevic, 2020; Mathew et al., 2017; Mathew et al., 2018; McGuire et al., 2021; Undurraga et al., 2020). It is considered a robust response, indicative of a perceived change in an ongoing stimulus (Martin et al., 2008). Ross suggested that the envelope ITD processing at the subcortical level requires stimulus phase locking and the few individuals who exhibited responses at higher frequencies (2000 and 4000 Hz) might rely on stimulus cues other than the envelope ITD. Since for SAM tones with low carrier frequencies (<2000 Hz), the responses evoked by envelope ITD are mainly delivered by the fine structure (phase locking), only SAM tones with a high carrier frequency (4000 Hz) were used in experiment 2. In addition, previous physiological studies showed that the highest modulation rates to which envelope following responses are elicited also decreases with age (Purcell et al., 2004). Thus, multiple modulation rates were used to gauge the effect of modulation rate on responses to prepare for potential future applications.

To further simulate the CI performance in NH listeners, the envelope ITD sensitivity of bandpass-filtered pulse trains at different pulse rates was tested in experiment 3. Such bandpass-filtered pulse trains have been used to simulate CI performance in different studies (e.g., Carlyon et al., 2008; Hu et al., 2017; Hu et al., 2022; Laback et al., 2004), and were expected to evoke larger responses than the SAM tones (Hu et al., 2022) used in experiment 2.

Based on the findings of Ross (2018) for the 4000 Hz carrier frequency condition, we were cautious about whether the perceptually perceived envelope ITD changes are sufficiently salient to evoke strong ACC responses, especially in a clinical paradigm and with a simpler 'one-for-all' EEG data analysis pipeline. However, with the multi-information recording and analyzing techniques, even when the ACC responses evoked by the envelope ITD changes are not detectable, useful information from other types of responses is still expected, e.g., onset, offset CAEPs, and the ASSRs at multiple modulation rates. To our knowledge, this was the first EEG study to simultaneously obtain multiple cortical and subcortical responses at different rates to record the upper-frequency limits of the fine structure ITD and envelope ITD sensitivity from the same participants.

## Methods

### *Participants*

Twelve NH participants (S1 - S12: 6 males and 6 females aged 21- 42 years old, with a mean age of 27.2 years) took part in the psychoacoustic experiment, while only the first eight (S1 - S8: 5 males and 3 females aged 21 - 35 years old, with a mean age of 26.4 years) participated in the EEG experiments. None of the participants had a history of neurological, psychiatric, or otological disorders. All had audiometric thresholds of 20-dB hearing level or better at octave frequencies between 125 Hz and 8 kHz. The participants attended two appointments on separate days. The first appointment, lasting less than 30 minutes, was for psychoacoustic experiments. Participants were encouraged to take breaks after each block (<10 minutes). The second appointment, excluding preparation, lasted approximately 120 minutes including breaks, was for the three independent EEG experiments (see the flow chart in Supplementary Figure 1). Participants provided voluntary written informed consent and were compensated with hourly pay for their participation, with the approval of the Ethics Committee of the University of Oldenburg (Drs.EK/2019/075).

### *Apparatus*

The Oldenburg Alternative Forced Choice (AFC) framework (Ewert, 2013), which is freely available, was used to control the experiments and present the stimuli to the participants in both psychoacoustic and EEG measurements. The stimuli were generated digitally using a personal computer running MATLAB (The Mathworks, Natick, MA, USA), then converted to analog form using a Fireface UC sound card (RME Audio, Haimhausen, Germany) with 24-bit resolution and a sampling rate of 48 kHz. In the psychoacoustic experiments, the stimuli were presented to the participants through Sennheiser HD580 headphones at 70 dB sound pressure level (SPL). The participants were seated in a double-walled soundproof booth and responded by clicking the virtual buttons displayed on a monitor.

During the EEG experiment, a Fireface UCX sound card was connected to Tucker Davis (Alachua, FL) HB7 headphone driver and presented to participants through ER-2 insert earphones (Etymotic Research, Elk Grove, IL, USA). The stimuli were calibrated to 75 dB SPL. Participants sat in a recliner and watched silent, subtitled movies in an electrically shielded soundproof booth while instructed to minimize movements. The EEG data were recorded differentially from Ag/AgCl electrodes via CURRY7 (Neuroscan) and a recording computer. This study used 14 out of 64-channel braincap (Easycap) electrodes (Figure 1A: 8, 10, 12, 14, 16, 31, 33, 42, 49, 55, 56, 59, 62, and 63). The FPz served as the ground and Cz as the physical reference. Four central anterior-posterior EEG recording electrodes (Figure 1B: channel 56 and 62, right and left mastoids; channel 49, Inion; channel 59, ~ 3.5cm below the Inion) were of primary interest for potential clinical applications (Hu and Dietz, 2015; Hu et al., 2015). Impedances were kept below 10 K$\Omega$ and checked after each run, adjusting if necessary. The scalp electrodes were connected to the 70-channel SynAmps RT amplifier system (Neuroscan) via monopolar input connectors. The voltage resolution was approximately 29.8 nV/LSB, and the recordings were filtered by an analog antialiasing lowpass filter with a corner frequency of 8 kHz, digitized with a 20 kHz via a 24-bit A/D converter.

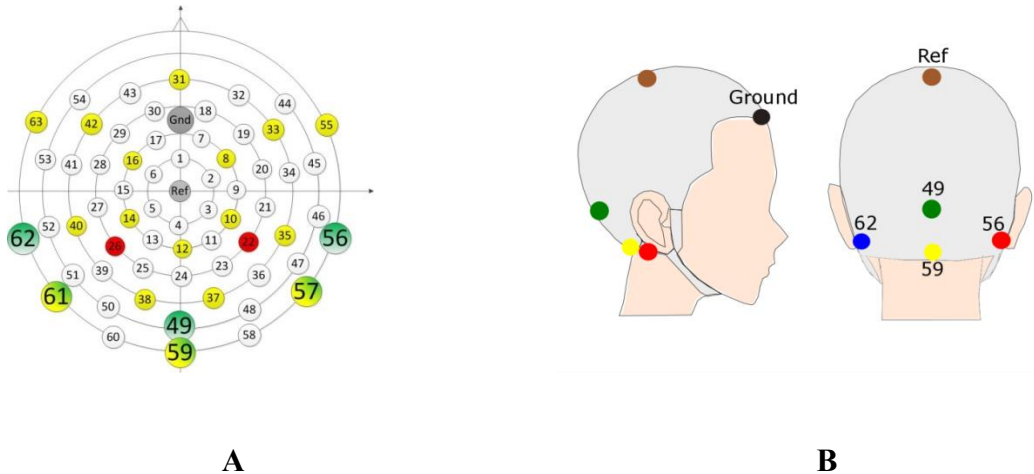

Figure 1 (color online): (A) the scalp channel locations and labels of the EEG cap (Hu et al 2015). The position on the central anterior-posterior line corresponds to a 10% electrode system, with electrode labels such as Fpz (31), Ref (Cz), and Iz(49), etc.Channels 49, 56, 59, and 62 were the channels of primary interest. (B) electrode location for clinic setup. Mastoid (left: blue,62; right: red, 56); Middle line (inion: green,49; yellow, 59); reference (Cz, brown); ground (forehead, black)

### *Psychoacoustic experiments*

Three lateralization experiments (5-7 minutes/experiment) were performed to determine the upper frequency limit of left/right discrimination abilities. The experiments used a two-up, one-down, two-alternative forced-choice procedure (2-AFC) to estimate the 71%-correct threshold on the psychometric function (Levitt, 1971). On each trial, two consecutive intervals were presented, separated by 500 ms. Each interval contained four consecutive 400-ms tones or filtered clicks, with 20-ms raised cosine rise/fall ramps, separated by 100 ms. One interval was randomly selected as the standard and had an ITD of the carrier (ITD$_{FS}$) or ITD of the envelope (ITD$_{ENV}$) of 0. The other interval, the target, had the same first and third tones as the standard, but the second and fourth tones had a non-zero ITD of the same magnitude as each other. During all three experiments, participants were asked to identify which of the two intervals contained a sequence that appeared to change within the head.

Experiment 1 was conducted to determine the upper limit of the carrier frequency ($f_{uplim\_c}$) for fine structure ITD sensitivity by applying an IPD of $\pi/2$ to the carrier frequency. The experiment utilized SAM tones with a fixed modulation frequency ($f_m$) of 40 Hz and an adaptive carrier frequency ($f_c$) ranging from 100 Hz to 4000 Hz. The carrier frequency was adjusted using adaptation factors of 1.4, 1.2, and 1.1, starting at 1000 Hz. Before the formal experiment, a brief training task was provided to familiarize the participants with the procedures. After eight reversals, the formal test was terminated and the threshold was calculated as the geometric mean of the last six reversal values. This procedure was adapted from the binaural TFS sensitivity test (TFS-AF) (Füllgrabe et al., 2017; Füllgrabe and Moore, 2017). The reason for its selection is that it has been validated in both young and old NH participants (Füllgrabe et al., 2017; Füllgrabe and Moore, 2017; Grose and Mamo, 2010; Moore et al., 2012; Ross et al., 2007a; Ross et al., 2007b; Thorup et al., 2016), and it is conceptually similar to the ACC paradigm used in the previous investigations (e.g., Grose and Mamo, 2010; Ross et al., 2007a; Ross et al., 2007b).

Experiment 2 was conducted to determine the upper limit of the modulation rate ($f_{\text{uplim}\_m}$) for envelope ITD sensitivity, with a fixed ITD of 500 μs (dichotic) or 0 (diotic) applied to the envelope. The $f_c$ was fixed at 4000 Hz, while the $f_m$ was adaptive. Experiment 3 was performed to determine the upper limit of the pulse rate ($f_{\text{uplim}\_pps}$) for interaural pulse time difference (IPTD) sensitivity, using an ITD of 500 μs applied to pulses and adjusting the pulse rate. Both experiments were similar to experiment 1, except that in experiments 2 and 3, the start $f_m$ or pulse rate was 100 Hz or 100 pps, with a minimum and maximum of 10 Hz or pps and 1500 Hz or pps, respectively. The adaptation factors were 80, 40, and 20. To accommodate individual differences, an upper limit for the adaptive $f_m$ was not set. This may have resulted in some participants perceiving lower sidebands of the modulation as audible for higher modulation rates (e.g., above ~350 Hz) (Kohlrausch et al., 2000), transforming the ITD change detection task into a disparity detection task not only based on envelope ITD cues. The decision not to set an upper limit was made to consider the possibility that participants may use other cues, which could also trigger ACC responses. However, caution should be taken when interpreting the upper modulation rate regarding the ITD sensitivity, especially the $f_{\text{uplim}\_m}$, where some participants might use other cues (e.g., the lower sideband) than ITD in the detection task.

***Stimuli***

*SAM tones*

In both experiments 1 and 2, SAM tones were generated digitally according to equation (1) (Bernstein and Trahiotis, 2012; Hu et al., 2022).

$$s(t) \,=\, a\,sin(2\pi f_c\,t)(1 - cos2\,\pi f_m t) \qquad (1)$$

Figure 2A, column 1, shows an example of a stimulus used in the psychoacoustic experiment 1 (SAM tones ITD$_{FS}$, $f_c$ = 1000 Hz and $f_m$ = 40 Hz). In this example, the first interval is the target (row 2), and the second interval is the standard (row 3). Both the psychoacoustic and EEG experiments employed IPD of 0 or $\pi/2$. EEG experiment 1 tested four carrier frequencies, $f_c$ = [200, 400, 800, 1600] Hz. Figure 2B, column 1, shows an example of a stimulus used in EEG experiment 1 (SAM tones ITD$_{FS}$, $f_c$ = 400 Hz and $f_m$ = 40 Hz), where each presentation lasted 8 seconds (s). The sequence included 2 s of the diotic stimulus (IPD = 0 in time window T1), followed by 2 s of the dichotic stimulus (IPD = $\pi/2$ in time window T2; T1$\rightarrow$T2 referred to as outward switching), then 2 s of the standard stimulus (IPD = 0 in time window T3; T2$\rightarrow$T3 referred to as inward switching), and 2 s of silence (in time window T4).

The stimuli used in the second psychoacoustic experiment (not shown in Figure 2A) were also SAM tones, but the ITD was applied to the envelope instead of the carrier. In EEG experiment 2, four modulation frequencies were tested $f_m$ = [40, 80, 160, 320] Hz. Figure 2B, column 2 shows an example of a stimulus used in this experiment (SAM tones ITD$_{ENV}$, $f_c$ = 4000 Hz and $f_m$ = 40 Hz). Like in EEG experiment 1, it consisted of 2 s of the diotic stimulus (ITD$_{ENV}$ = 0 in the time window

T1), followed by 2 s of the dichotic stimulus (ITD$_{ENV}$ = 500 µs in the time window T2; with an outward switching T1→T2), then again 2 s of the standard stimulus (ITD$_{ENV}$ = 0 in the time window T3; with an inward switching T2→T3), and 2 s of silence (in the time window T4). As Ross (2018) showed no detectable ACCs in most of their participants for the 4000 Hz SAM, the ACCs in experiment 2 might be smaller than in experiment 1 or absent.

Note that, as in (Kohlrausch et al., 2000), no precautions were taken to mask possible distortion products in both psychoacoustic and the EEG experiment 2. Nevertheless, with the modulation frequencies below 320 Hz in the EEG experiments, we believe that the main results summarized below are not affected by detection cues based on distortion products.

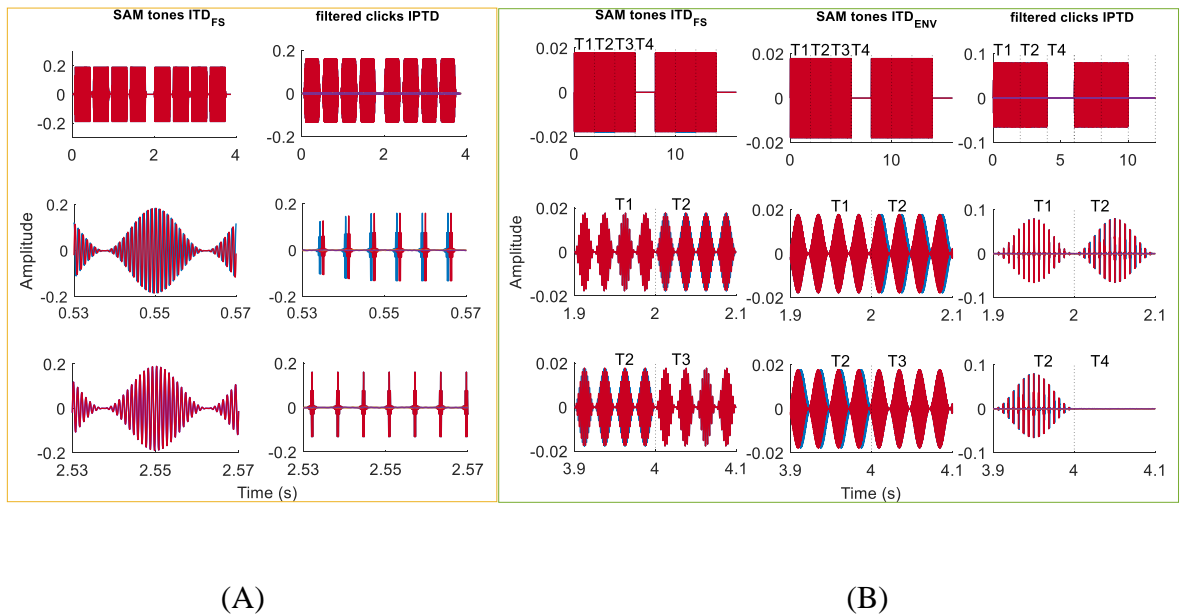

(A)                                      (B)

Figure 2 (color online): Exemplary stimuli used in the psychoacoustic (A) and EEG experiments (B). Two intervals of SAM stimuli (A, column 1, $f_m$ = 40 Hz, $f_c$ = 1000 Hz, IPDfs = π/2) and filtered clicks (A, column 2, pulse rate = 160 pps, $f_m$ = 2.5 Hz, IPTD = 0 or IPTD = 500 µs) in the psychoacoustic experiments. Two consecutive intervals were separated by 500 ms of silence. Each interval contains four consecutive 400-ms tones (including 20-ms cosine rise/fall ramps), separated by 100 ms. In these examples, the first interval is the target, where the first and third tones were the same as in the standard interval while the second and fourth tones differed in their IPD by π/2 or by IPTD of 500 µs. Two repeats of the SAM stimuli with ITD$_{FS}$ (B, column 1, $f_m$ = 40 Hz, $f_c$ = 400 Hz, IPD = π/2), SAM stimuli with ITD$_{ENV}$ (B, column 2, $f_m$ = 40 Hz, $f_c$ = 4000 Hz, ITD = 500 µs), and filtered clicks (B, column 3, pulse rate = 160 pps and $f_m$ =10 Hz, IPTD = 0 or IPTD = 500 µs) were used in three EEG experiments. The duration of each presentation is 8 s (experiments 1 and 2) or 6 s (experiment 3). It includes 2 s of the diotic stimuli (time window T1), 2 s of the dichotic stimuli (time window T2; T1→T2 named outward switching), 2 s of the standard stimuli (time window T3; T2→T3 named inward switching; for experiment 1 and 2), and 2 s of silence (time window T4).

*Filtered clicks*

In experiment 3, filtered clicks generated as in (Hu et al., 2017; Hu et al., 2022) were used to simulate the signal delivered to CI users. The pulse train was band-limited to 3-5 kHz with a center frequency of $f_c = 4$ kHz. These band-limited pulse trains $p(t)$ were then sinusoidally amplitude-modulated using formula (2).

$$s(t) = p(t) * [1 - cos(2\pi * f_m * t)] \qquad (2)$$

The $f_m$ in the psychoacoustic was 2.5 Hz (reciprocal of the duration of consecutive filtered clicks, i.e. 1/0.4s), while it was 10 Hz in EEG experiments. This type of SAM ensures that stimuli start at the trough of the modulation.

Figure 2 A (column 2) and B (column 3) show an example of the stimuli used in the psychoacoustic (filtered clicks IPTD, pulse rate = 160 pps and $f_m$= 2.5 Hz, IPTD = 0 or IPTD = 500 µs) and EEG experiments (filtered clicks IPTD, pulse rate = 160 pps and $f_m$= 10 Hz, IPTD = 0 or IPTD = 500 µs), respectively. In the EEG experiment, four fixed pulse rates of [40, 80, 160, 320] pps were used. The duration of each presentation is 6 s, which includes 2 s of the diotic stimulus, followed by 2 s of the dichotic stimulus (with a transition from T1➔T2, referred as outward switching), and 2 s of silence (in time window T4, with a T2➔T3 inward switching).

In both the psychoacoustics and EEG experiment 3, a low-pass noise, uncorrelated between the ears, was added to the filtered clicks to conceal potential distortion products. The low-pass noise was created by generating broadband noise in the time domain, converting it to the frequency domain, and setting the power of all components above 1000 Hz to zero. The noise was then manipulated to have a flat spectrum up to 200 Hz with a decreasing spectral density of 3 dB/octave above 200 Hz. It was further filtered with a 5th-order, lowpass filter with a cut-off frequency of 1000 Hz (Hu et al., 2017), and gated with 50-ms raised cosine ramps. The test stimulus was centered within the noise presentation, which was presented at 40 dB SPL.

We chose 4000 Hz instead of a higher carrier frequency such as 8000 Hz for several reasons: Firstly, Previous studies have shown that the upper modulation rate is lower for stimuli centered at 8000 Hz compared to those centered at 4000 Hz (Bernstein and Trahiotis, 2013). Since only 2 out of 14 participants in Ross (2018) showed significant responses at 4000 Hz, we would expect similar or even smaller responses at 8000 Hz. Secondly, as the aging population is one target group for future studies, high-frequency hearing loss may make a higher frequency less optimal. Lastly, although it is not a critical factor, 8000 Hz is less pleasant to listen to than 4000 Hz.

### *EEG experiments*

For each participant, 60 repetitions (4 runs/$f_c$ ×15 repeats/run) were carried out in total per test condition during the EEG appointment. The appointment had a duration of approximately 35 minutes each for experiment 1 (with four carrier frequencies $f_c$ of [200, 400, 800, 1600] Hz) and experiment 2 (with four modulation frequencies $f_m$ of [40, 80, 160, 320] Hz), and a duration of approximately 25 minutes for experiment 3 (with four pulse rates of [40, 80, 160, 320] pps). Each run consisted of 15 repetitions per condition, which were separated into 3 blocks, each block comprising 5 repetitions of the same 8 s (Figure 2B, columns 1 and 2), or 6 s (Figure 2B, column 3) stimuli. This resulted in 12 blocks per run (3 blocks/condition × 4 conditions/run). The blocks were randomized in each run. During the recording, any ongoing artifact rejection was turned off, and filtering, artifact analysis, and averaging were done offline.

### *EEG data analysis*

For the EEG results, continuous EEG data collected from each participant were segmented into epochs over a window of 8.2 s or 6.2 s, including a pre-stimulus duration of 200 ms. The data were digitally filtered using a two-order Butterworth band-pass filter with a frequency range of 0.1–1000 Hz following the segmentation. The baseline was corrected by the mean amplitude of the last 1 s after the stimulus offset time window, and epochs with voltages exceeded ± 200 μV were rejected from further analysis. Then, the EEG data was averaged separately for each condition in each participant. As Cz is a commonly used channel for cortical responses, the recordings in this study were re-referenced to the average of the four clinical recording channels (Figure 1B, 56, 62, 49, and 59).

To obtain the transient response in the time domain, the responses were filtered further using a two-order Butterworth band-pass filter with a commonly used frequency range of 0.1–30 Hz (e.g., Calcus et al., 2022; Papesh et al., 2017). The peak amplitudes and latencies of the auditory-evoked P1, N1, and P2 for each participant, as well as the average response across all participants, were automatically identified within a fixed time windows of 10-85, 85-160, and 160-300 ms after the onset, change, or offset of the stimuli, respectively. To obtain the ASSRs in the frequency domain, the filtered data between 0.1-1000 Hz was used. ASSRs within different time windows (T1, T2, T3, T4, and T1234 in EEG experiment 1& 2, T1, T2, T4, and T124 in EEG experiment 3) were obtained, where T1234 and T124 refer to the whole duration of each stimulus (8s or 6s).

To explore the time-frequency characteristics of the evoked responses, a complex Morlet wavelet $\omega$ defined as equation 3 (Cohen, 2018) was used for visualization, where $f$ is the frequency in Hz, $t = -1 : \frac{1}{f_s} : 1$ is the time in seconds, $f_s$ is the sampling rate, and $i = \sqrt{-1}$. $\sigma$ is the width of Gaussian, and $n$ is the number of cycles. Normally n is between 2 to 15 for neuro-electrical signals with frequencies between 2 Hz and 80 Hz. In this study, $n = 6$ unless otherwise stated.

$$\omega = e^{2i\pi f t} e^{-\frac{t^2}{2\sigma^2}}, \ \ \sigma = n/(2\pi f) \quad (3)$$

*Statistical analysis*

Statistical analyses were conducted using IBM SPSS (version 27, IBM Corp., Armonk, NY). To assess the effect of stimulus type (i.e. carrier frequency in experiment 1, modulation rate in experiment 2, and pulse rates in experiment 3) and response type (onset CAEP; outward change response ACC1; inward change response ACC2; offset CAEP) on the amplitude (e.g., N1P2, ASSR) and latency (e.g., N1), a general linear model repeated-measures (GLMrm) analysis was performed. If the sphericity assumption was violated, a Greenhouse-Geisser correction was applied. In some cases, pairwise post-hoc comparison $t$-tests with Bonferroni correction were performed for further analysis. The reported $p$-values are adjusted using Bonferroni correction provided by SPSS, with the number of comparisons adjusted based on the number of levels in the assessed factor. Unless stated otherwise, the significance threshold reported in the results is $p < 0.05$.

To examine the correlation between behavior test results ($f_{uplim\_c}$, $f_{uplim\_m}$, or $f_{uplim\_pps}$), and EEG results (N1P2 amplitude of various CAEPs and ASSR amplitude at the respective modulation frequency), Pearson's correlation coefficient r was calculated. The reported correlations generally had a p < 0.05, unless stated otherwise. We did not account for the familywise error rate when dealing with multiple comparisons but this will be borne in mind in the interpretations.

**Results**

*Psychoacoustic results*

Figure 3 shows the violin plots of the $f_{uplim\_c}$, $f_{uplim\_m}$, and $f_{uplim\_pps}$ from three psychoacoustic experiments. The violin plots (Hintze and Nelson, 1998) were generated using freely available Matlab code (https://github.com/bastibe/Violinplot-Matlab). The original box plot shape is included as a grey box in the center of the violin. Figure 3 depicts the individual data of the 12 participants as solid blue dots that have been randomly jittered from the center. The corresponding density curves have been constructed around each center line. If the participant couldn't do the task, the value was set to 0.123456. Figure 3 indicates that the upper limits vary across participants. The mean and standard deviation of $f_{uplim\_c}$ is 1393 ± 284 Hz, which is in the range of previously reported values (Brughera et al., 2013; Füllgrabe and Moore, 2017, 2018; Grose and Mamo, 2010; Hopkins and Moore, 2010, 2011; Papesh et al., 2017; Ross et al., 2007a; Ross et al., 2007b).

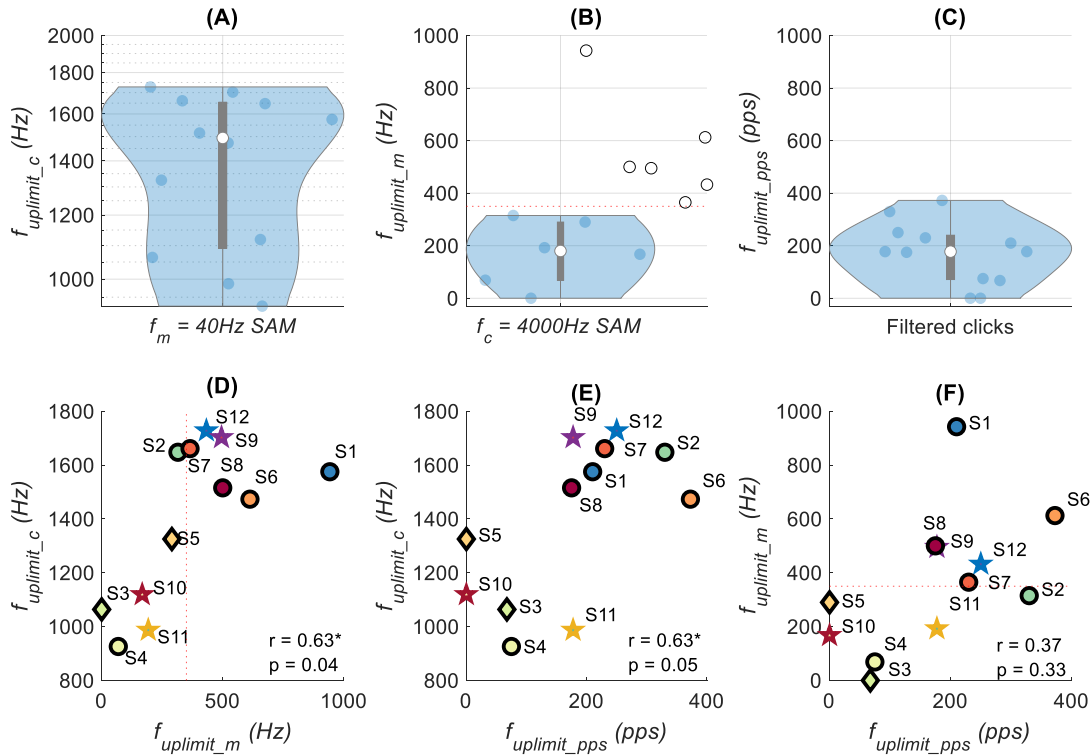

Figure 3 the top panels of the figure show violin plots of the upper limit frequency $f_c$ obtained from the psychoacoustic experiment for each participant, represented by solid dots in each violin plot. The bottom panels display the correlation between the three upper limit frequencies. Participants S9-S12 (represented by pentagram symbols) were unable to attend the EEG experiment, while S3 couldn't achieve $f_{uplim\_m}$ and S5 couldn't achieve $f_{uplim\_pps}$, represented by diamond symbols. The dotted red lines in panels B, D, and F indicate the boundary of 350 Hz.

The top middle panel of Figure 3 shows the $f_{uplim\_m}$. Without setting a limit in the adaptive procedure, some participants reached $f_{uplim\_m}$ above 350 Hz. Participant S1 even reached 980 Hz. This was expected because the task may become easier again for some participants if they are able to use spectrally resolve sidebands at modulation rates above a certain frequency (e.g., ~ 350 Hz, the red dotted horizontal line) (Kohlrausch et al., 2000). This phenomenon may be more prominent in the disparity detection test procedure used in this study, compared to the classical left/right discrimination tasks. To avoid misleading interpretation, the mean and standard deviation of $f_{uplim\_m}$ were calculated after excluding data from participants who couldn't complete the task (S3) and those with $f_{uplim\_m}$ above 350 Hz (6 data points as indicated by the empty circles in the upper middle panel of Figure 4, which may be a result of the resolved sidebands). The resulting mean and standard deviation were 207 ±99 Hz. Some caution is necessary when interpreting the correlation between the $f_{uplim\_m}$ and other experimental results. However, the same issue was not apparent for the EEG results shown in Section 3.2, because the maximum modulation frequency tested in the EEG experiment was limited to 320 Hz. The mean and standard deviation of $f_{uplim\_pps}$ for filtered clicks, after excluding S5 and S10, were 207 ± 97 pps. The Pearson correlation coefficients between the three upper limits are as follows: between $f_{uplim\_c}$ and $f_{uplim\_m}$ (exclude S3), r = 0.63, p = 0.04; between $f_{uplim\_c}$ and $f_{uplim\_pps}$ (excluded S5 and S10), r = 0.63, p = 0.05; and between $f_{uplim\_m}$ and $f_{uplim\_pps}$ (excluded S3, S5, and S10), r = 0.37, p = 0.33.

S3 and S5 were unable to perform the corresponding experiments. However, both detected changes when presented with 100 Hz SAM tones and 100 pps filtered clicks with 500 μs ITD. It was speculated that this was mainly due to the large initial adaptive stepsize and their difficulty in focusing during the first reversal. S8 reported that he occasionally experienced mild tinnitus. Despite this, he was included in the EEG experiment as his audiometry results were within normal hearing range and his lateralization performance was above average. Regrettably, participants S9-S12 were unable to attend the EEG experiments due to reasons relating to the COVID-19 pandemic.

### EEG results

#### Time domain (CAEPs)

Figure 4 A-C show the individual (gray) and average EEG responses for experiments 1-3 in the time domain. The same processing procedure and automatic peak-detection method were applied to all conditions. Each figure presents the four test conditions using different colors. The P1, N1, and P2 peaks of the average responses are marked with triangle symbols in each of the test conditions in panels 1-4. For easy comparison, the averaged data from panels 1-4 are overlaid in panel 5.

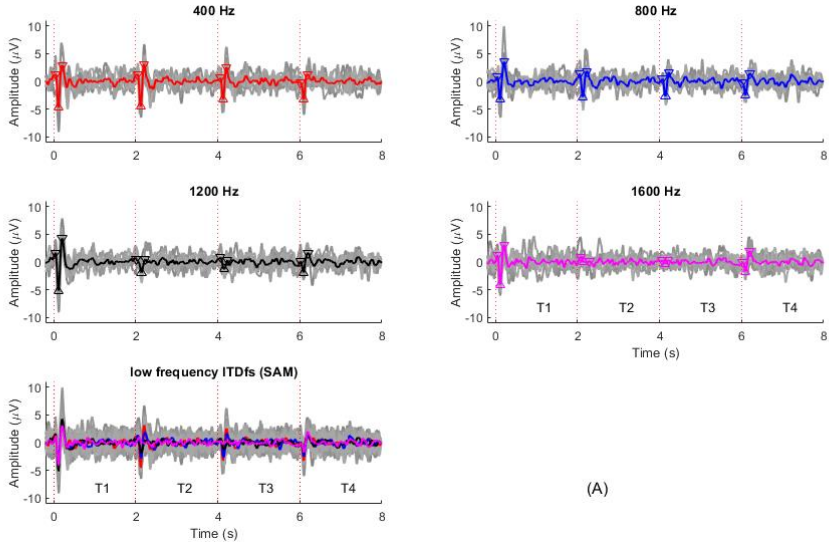

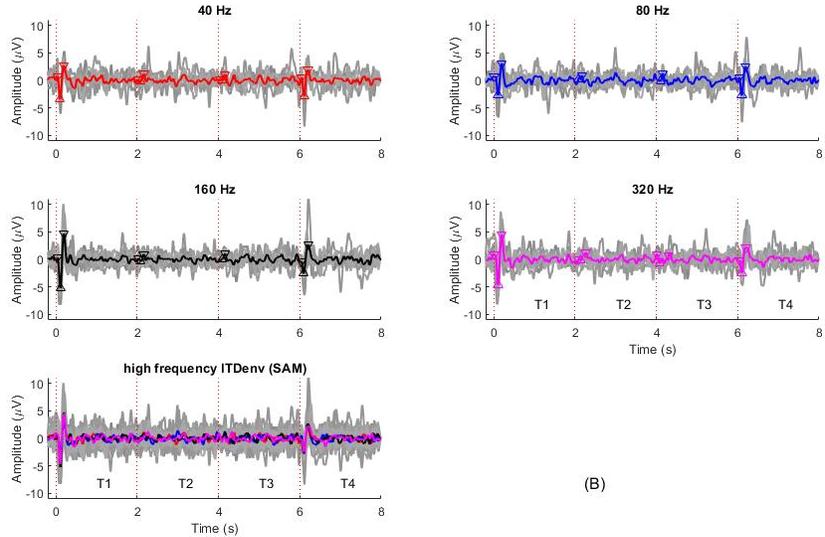

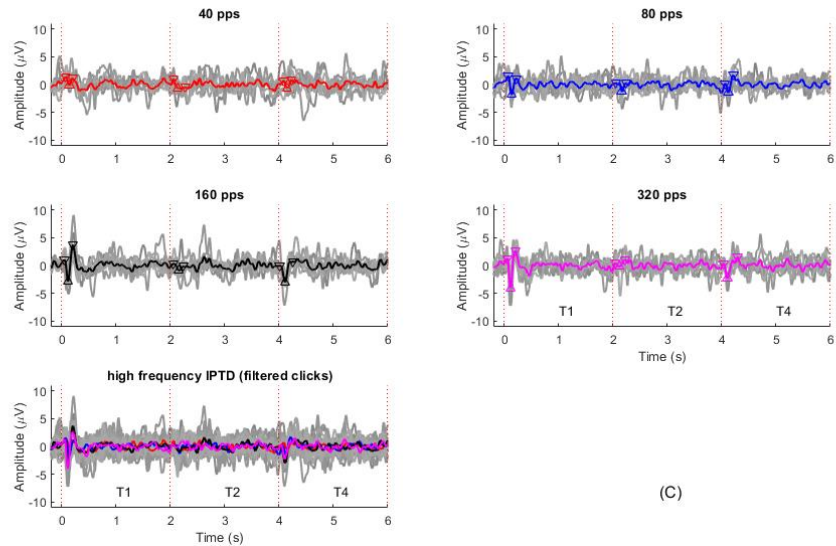

Figure 4 the average response in the time domain. The red, blue, black, and pink waveforms are the overall average CAEPs across participants for the four test conditions. The gray curves are the average CAEPs of each participant. The dotted red vertical lines are the start time of onset, ACC1, ACC2, and offset. T1, T2, T3, and T4 are the corresponding time duration. The four conditions are $f_c$ = [200, 400, 800, 1600] Hz in (A), $f_m$ = [40, 80, 160, 320] Hz in (B) and pulse rate = [40, 80, 160, 320] pps in (C).

Figure 5 re-plots some results from Figure 4. The upper panel shows the average responses evoked by 40 Hz SAM tones with $\text{ITD}_{\text{FS}}$ changes ($f_c$ = 400 Hz, named, fc400ITDfs, red) and by 40, 80, 160, and 320 Hz SAM tones with $f_c$ = 4000 Hz and $\text{ITD}_{\text{ENV}}$ changes. The lower panel shows the responses evoked by 40, 80, 160, and 320 pps filtered clicks with IPTD changes, in addition to the fc400ITDfs (red). By overlaying these curves, some ACC-like responses were able to be identified within the same ACC time duration as fc400ITDfs (after 2 and 4 s, red curve). However, the ACC responses evoked by envelope ITD changes were much smaller than those evoked by $\text{ITD}_{\text{FS}}$ and close to the noise floor for both 4000 Hz SAM tones (upper panel) and filtered clicks (lower panel). The automatically marked ACC peaks mostly fall within the same range as the noise within the detection windows in experiments 2 and 3, and the 1600 Hz condition of experiment 1. Thus caution should be taken when interpreting the ACC peaks in these cases.

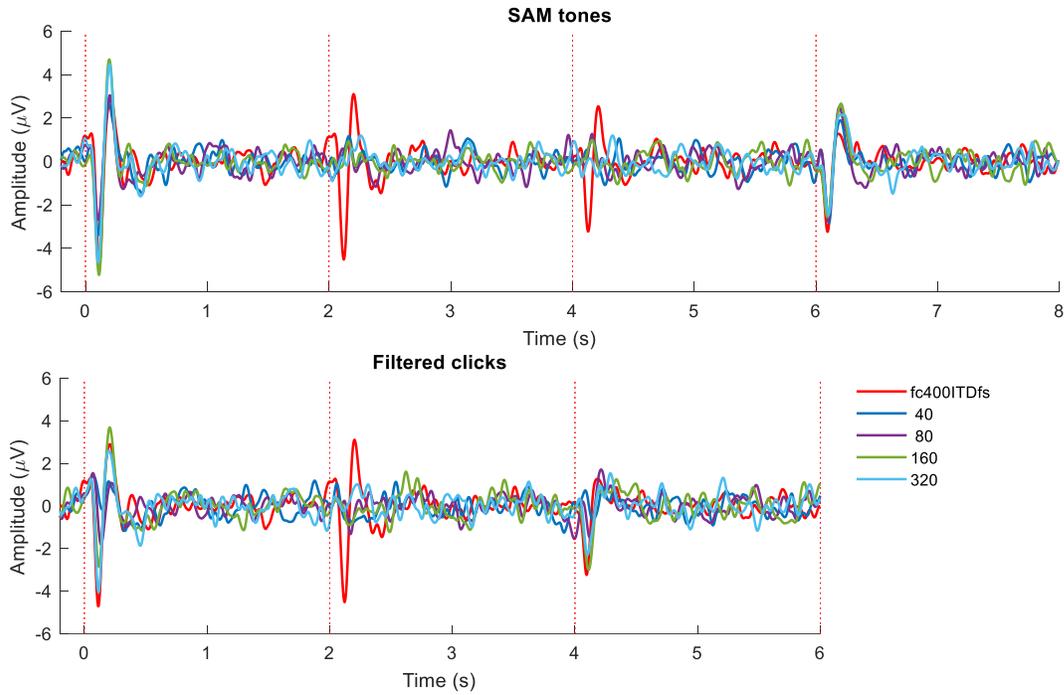

Figure 5 Overlay of average EEG responses for high-frequency SAM tones and filtered clicks with modulation rate of 40,, 80, 160, and 320 Hz or pps. The red curve represents the average response for 400 Hz low-frequency SAM tones.

By observing the waveform morphology of the EEG responses shown in Figure 4 - Figure 5, and considering the information redundancy among P1, N1, and P2, in the statistical analysis, mainly the results of N1P2 amplitude (the amplitude difference between P2 and N1, i.e., P2-N1, in µV) and the N1 latency were reported. Figure 6 (A-C) shows the violin plots of the automatically detected N1P2 amplitude and N1 latency of experiment 1-3. The pair-wise Bonferroni corrected *t*-tests with p<0.05 were marked with '*' symbols.

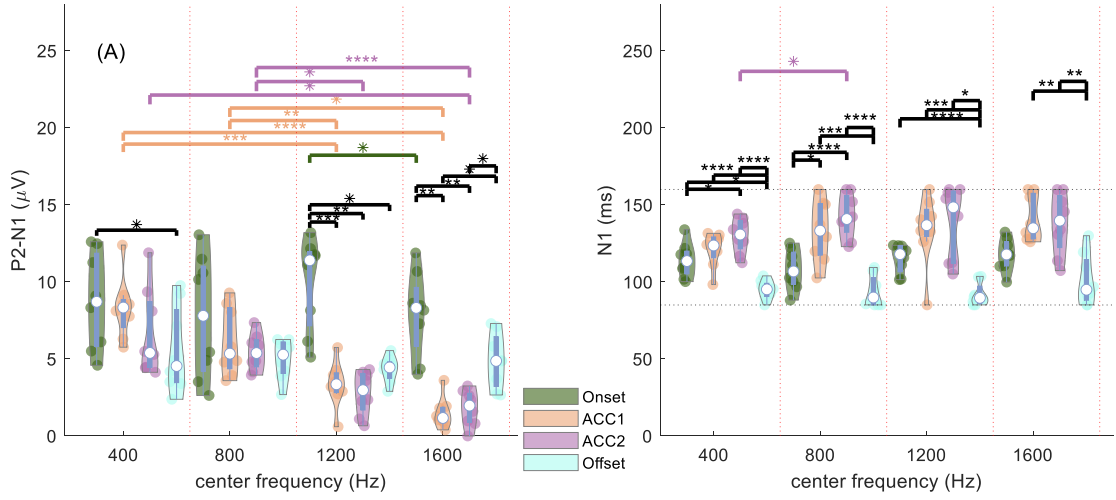

(A)

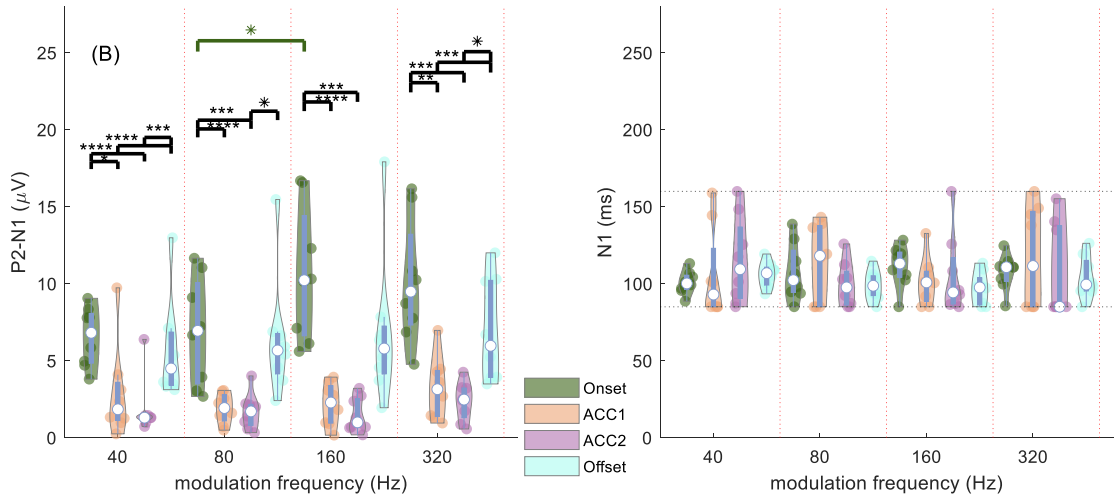

(B)

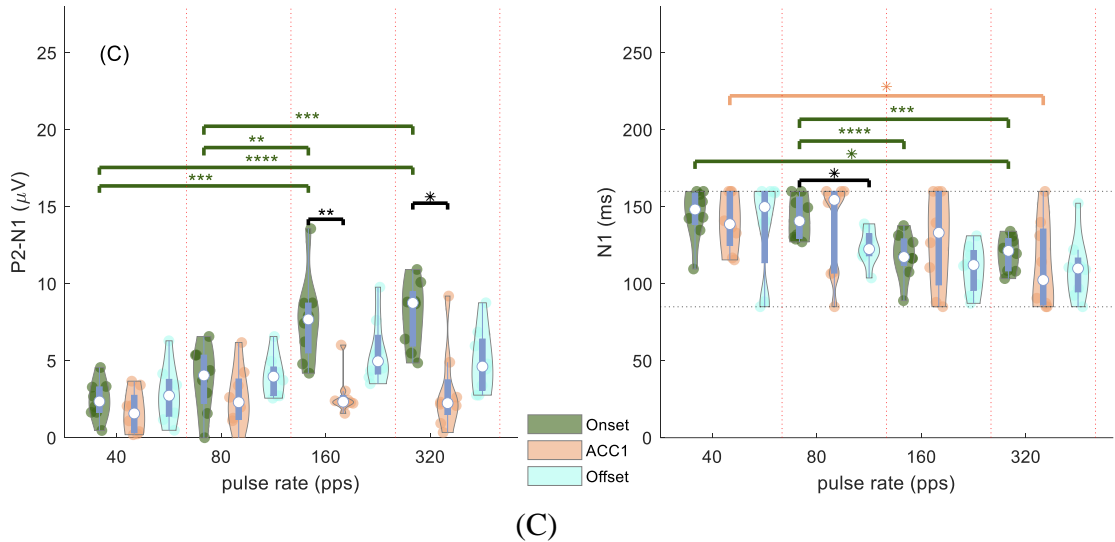

(C)

Figure 6 violin plots of the N1P2 amplitude (left column, in μV) and the N1 latency (right column, in ms) of four response types (color coded) under four test conditions (separated by the vertical red dotted lines). (A) the four conditions are $f_c$ = [200, 400, 800, 1600] Hz. (B) the four conditions are $f_m$ = [40, 80, 160, 320] Hz. (C) the four conditions are pulse rate = [40, 80, 160, 320] pps (A) carrier frequencies (400, 800, 1200, 1600 Hz, separated by the vertical red dotted lines). The solid dots within each violin plot are individual data from each participant. Conditions marked with asterisks (*; **; ***; ****) are significantly different (p<0.05; p<0.01; p<0.005; p<0.001). Black lines indicate comparisons between different response types within each frequency group, while different colors represent comparisons within a specific response type across frequencies.

- **CAEPs of experiment 1**

Regarding experiment 1 (ITD$_{FS}$, Figure 4 A and Figure 6 A), the amplitude and latency of the offset responses (aqua) were relatively consistent across different carrier frequencies and the N1 latency of the offset responses was generally shorter compared to the onset and ACC responses.

The N1P2 amplitude was significantly affected by carrier frequency ($f_c$), response type, and their interaction according to GLMrm (p<0.005). The mean amplitude was 7.318/6.063/5.094/4.023 μV for 400/800/1200/1600 Hz, respectively. There were no significant differences between 400, 800, and 1200 Hz, but the N1P2 amplitude for the 1600 Hz was significantly smaller than the other carrier frequencies. For the offset CAEPs, there were no significant differences between carrier frequencies. For most of the onset CAEPs, the differences were not significant except that the N1P2 amplitude of 1200 Hz was slightly larger than that of 1600 Hz (p = 0.048). For ACC1 (outward) responses, the N1P2 amplitudes of 400 Hz and 800 Hz were significantly larger than the 1200 and 1600 Hz, but there were no significant differences between 400 and 800 Hz, and between 1200 Hz and 1600 Hz.

For ACC2 (inward) responses, the N1P2 amplitude of 800 Hz was significant larger than that of 1200 and 1600 Hz, and the N1P2 amplitude of 400 Hz was significantly larger than that of 1600 Hz.

The mean N1P2 amplitudes were 8.577/4.816/4.172/4.932 μV for onset/ACC1/ACC2/offset responses, respectively. The onset CAEPs were significantly larger than the ACC1, ACC2, and offset CAEPs. However, there were no significant differences between the three latter types. Pairwise comparisons within each $f_c$ showed that the onset CAEPs were significantly larger than the offset CAEPs only for 400 and 1200 Hz. There were no significant differences between ACC1 and ACC2 for all carrier frequencies. Significant correlations were observed between the onset N1P2 amplitudes of most carrier frequencies, except for 400 vs 1200 Hz, and 400 vs 1600 Hz.

The mean N1 latency was 114/132/137/95 ms for onset/ACC1/ACC2/offset, respectively. The GLMrm analysis showed a significant effect of response type (p<0.001), but no significant effect of $f_c$ and their interaction. Pairwise comparisons showed significant differences between most response types (p<0.01), except between ACC1 and ACC2. The N1 latency of ACC responses was significantly larger than the onset response, while the offset response had the shortest latency and was significantly smaller than the other response types.

In summary, the results from experiment 1 were generally consistent with Ross et al. (2007b). For example, the mean P1, N1, and P2 amplitudes of ACC were smaller than those of the onset response: P1, 1.684/1.405/1.005/0.248 μV; N1, 3.324/-1.299/-1.019/-2.146 μV; P2, 3.946/2.128/1.862/2.379 μV for onset/acc1/acc2/offset. The ACC latencies were delayed compared with the corresponding onset and offset ones: P1, 42/46/57/27 ms; N1, 114/132/137/95 ms; P2, 211/227/240/213 ms for the onset/ACC1/ACC2/offset. The mean latencies of both P1 and N1 were in the same range but slightly smaller than Ross et al. (2007a). The latencies of ITD$_{FS}$ change evoked ACC1 and ACC2 that were longer than the onset, and the differences were smaller than Ross et al. (2007a). Consistent with (Ross, 2018), there was a tendency for larger responses to outward IPD changes (ACC1) than inward changes (ACC2) for the lower carrier frequencies, however, it was not significant here (p>0.5).

- **CAEPs of experiment 2**

In experiment 2 (ITD$_{ENV}$), as demonstrated in Figure 4B and Figure 6B, similar to experiment 1, there were clear onset and offset responses in all four test conditions. The onset N1P2 amplitude was comparatively larger than the offset responses, but the difference between the onset and offset CAEPs was smaller compared to those shown in Figure 4A and Figure 6A. Consistent with the findings of Ross (2018), the N1P2 amplitudes of both onset and offset CAEPs were larger than the ACC responses, due to the tiny (close to the noise floor) or absence of ACC responses.

Regarding the N1P2 amplitude in experiment 2, a GLMrm analysis revealed significant effects of $f_m$, response type, and their interaction. The mean amplitude was 4.249/4.228/5.258/5.665 μV for $f_m$ of 40/80/160/320 Hz, respectively. A significant difference between $f_m$ was only observed for 80 Hz vs 160 Hz. Consistent with this, pairwise comparisons within each response type only showed a just significant smaller onset N1P2 amplitude in 80 Hz condition compared to the 160 Hz condition (p = 0.048). The mean amplitude was 8.547/2.543/1.853/6.458 μV for onset/ACC1/ACC2/offset, respectively. Both onset and offset CAEPs were larger than the ACC responses. There were no significant differences between ACC1 and ACC2, and between onset and offset. Within each $f_m$, pairwise comparisons also showed no significant difference between onset and offset CAEPs, and between ACC1 and ACC2 responses (near noise floor). The onset and offset CAEPs were significantly larger than ACC responses, except for the comparison between ACC1 and offset for $f_m$ = 80 Hz, and between ACC1 and offset, and ACC2 and offset for $f_m$ = 160 Hz. Significant correlations were observed among modulation frequencies for all onset N1P2 amplitudes and for most offset N1P2 amplitudes, except for 320 vs 40, and 320 vs 160 Hz. The offset CAEPs were more correlated with the ACC responses than with the onset responses, mainly due to their small amplitudes.

For N1 latency, the GLMrm analysis showed no significant effect of either $f_m$ or response type. The mean latency was 107/109/107/101 ms for onset/ACC1/ACC2/offset, and 107/105/104/109 ms for 40/80/160/320 Hz.

- **CAEPs of experiment 3**

Regarding the filtered clicks (Figure 4C and Figure 6C), there were no ACC2 responses (for inward IPTD changes) recorded in experiment 3. Overall, the N1P2 amplitude of both onset and offset responses increased with increasing pulse rates. Similarly to experiment 2, the ACC1 responses were either small (near the noise floor) or absent.

For N1P2 amplitude, GLMrm showed a significant effect of pulse rate, response type, and their interactions (p<0.01). The mean amplitude was 2.31/3.43/5.36/5.35 μV for 40/80/160/320 pps, respectively. There were no significant differences between pulse rates of 40 and 80 pps, and between 160 and 320 pps. Within each response type, pairwise comparisons showed no significant differences between pulse rates for both ACC1 and offset responses. For the onset CAEPs, there were significant differences between most pulse rates (p <0.01), except for conditions of 40 vs 80 pps, and 160 vs 320 pps. The mean amplitude was 5.48/2.52/4.34 μV for onset/ACC1/offset, and only the difference between onset and ACC1 responses was significant. Within each pulse rate, pairwise comparisons showed no significant differences between response types for most pulse rates, except that for 160 pps and 320 pps, there was a significantly larger onset N1P2 amplitude than the offset one (p=0.009, and p = 0.014). There were no correlations between N1P2 amplitudes of different pulse rates for both onset and offset responses.

For N1 latency, GLMrm revealed a significant effect of pulse rate, but not of response types or their interactions. The mean latency was 141/134/119/114 ms for 40/80/160/320 pps, respectively. The N1 latency was significantly shorter for 320 pps compared to 40 pps and 80 pps, and for 160 pps compared to 80 pps. Within each response type, pairwise comparison showed significant N1 latency differences only for 320 pps vs 40 pps, 320 pps vs 80, and 80 pps vs 160 pps for the onset CAEPs, and between 40 pps and 320 pps (p = 0.044) for ACC1. The mean latency was 131/129/120 ms for onset/ACC1/offset responses with no significant differences between them. Within each pulse rate, there were nearly no significant differences between the three response types, except that the onset N1 latency was significantly larger than the offset one (p = 0.011) for the 80 pps.

*Frequency domain (ASSRs)*

Figure 7 shows the average ASSRs across participants within the analysis window of T1234 or T124. Within each panel, the colored curves in the shaded area are the average ASSRs for each individual. The red, blue, black, and pink spectra are the overall average ASSRs across participants for different test conditions: (A) $f_c$ = [200, 400, 800, 1600] Hz; (B) $f_m$ = [40, 80, 160, 320] Hz; (C) pulse rate = [40, 80, 160, 320] pps. The numbers with the same colors in the figure represent the ASSR amplitudes at the modulation frequency. The right panel displays the violin plots of the ASSR amplitude at the modulation frequency, within a duration of 8s (T1234) for the SAM tones or 6s (T124) for the filtered clicks.

For experiment 1 (Figure 7A), the group mean 40-Hz ASSR amplitudes were 0.187/0.173/0.152/0.142 μV for 400/800/1200/1600 Hz, respectively. The amplitude of the 40-Hz ASSR decrease gradually with increasing carrier frequency for the SAM-type stimuli, as previously reported by Ross (Ross, 2018). However, these differences are not statistically significant. The 40-Hz ASSR of SAM tones with different carrier frequencies were correlated, except for the comparison between 800 Hz and 1600 Hz. Additionally, the 40-Hz ASSR was correlated with the ACC2-N1P2 amplitude of 800 Hz. There was no correlation between the $f_{\text{uplim\_}c}$ and N1P2, and between the $f_{\text{uplim\_}c}$ and 40-Hz ASSR amplitude.

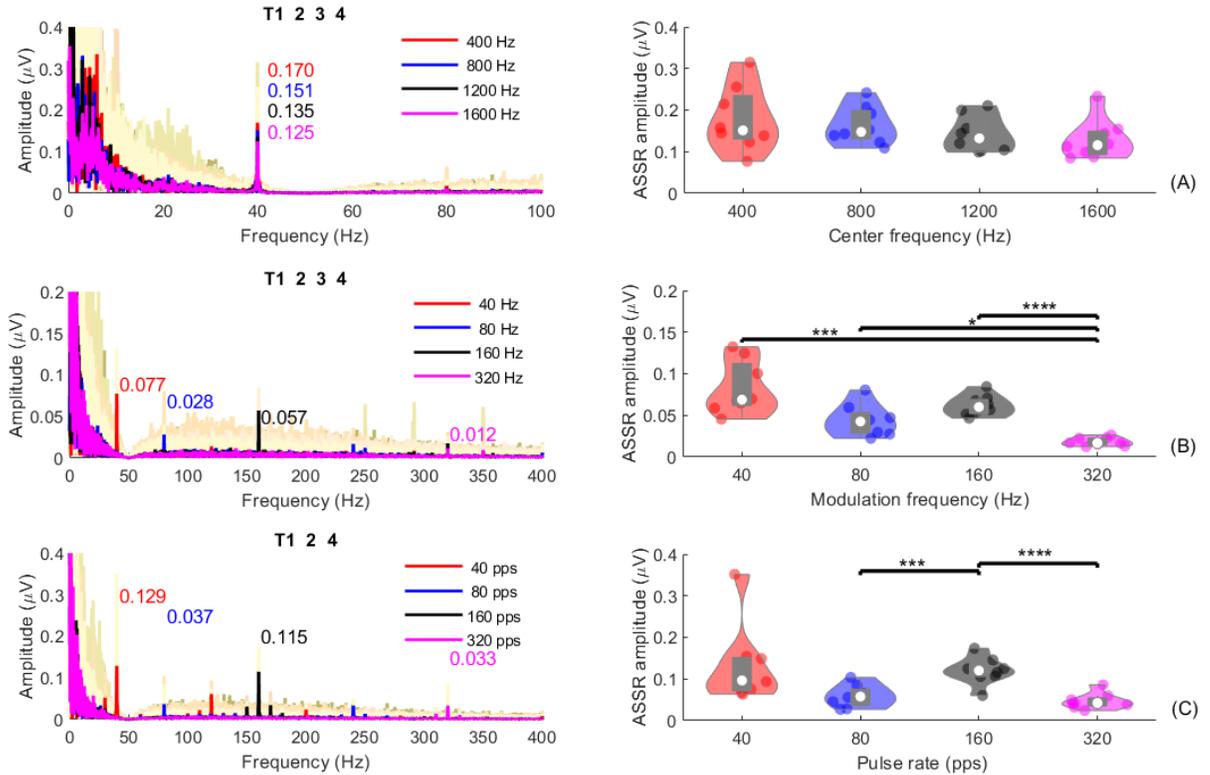

Figure 7 the individual and group average ASSRs in the frequency domain, with analysis time window T1234, or T124. The group average ASSRs across participants are depicted in red, blue, black, and pink for different test conditions: (A) $f_c$ = [200, 400, 800, 1600] Hz; (B) $f_m$ = [40, 80, 160, 320] Hz; (C) pulse rate = [40, 80, 160, 320] pps. The cream colored background curves are the average ASSRs for each individual. The numbers in different colors indicate the ASSR values at the modulation rate for different test conditions. The violin plots on the right display the corresponding ASSR amplitudes within an analysis window of 8s (T1234) for SAM tones or 6s (T124) for filtered clicks. The solid dots in each violin plot represent individual ASSRs at the corresponding $f_m$ or pulse rate for each participant. Conditions marked with asterisks (*; ***; ****) are significantly different (p<0.05; p<0.005; p<0.001).

Figure 7B and C show the ASSRs at different modulation rates (40, 80, 160, and 320 Hz). The numbers shown in different colors are the average ASSR values for each modulation rate. In general, the ASSRs for high carrier frequency stimuli (Figure 7B, with a smaller y-axis scale) were smaller than those for low-frequency (<=1600 Hz) SAM tones (Figure 7A), and the ASSRs for filtered clicks (Figure 7C) were larger than those for high-frequency SAM tones. In both types of high-frequency stimuli, the order was 40 Hz ASSR > 160 Hz ASSR > 80 Hz ASSR > 320 Hz ASSR. The filtered clicks elicited larger ASSRs, as suggested by Hu et al (2022).

The mean overall amplitudes of the ASSRs in experiment 2 (Figure 7B) were 0.093, 0.058, 0.071, and 0.022 µV for 40, 80, 160, and 320 Hz, respectively. The 320 Hz ASSR was significantly smaller than the others. There were no significant correlations between the ASSRs of different modulation frequencies, except for 160 Hz and 320 Hz. No significant correlation was found between the upper $f_m$ limit and the N1P2 amplitude or the ASSR amplitude. However, as noted in section 3.1, caution should be taken when interpreting any correlations with $f_{\text{uplim}\_m}$.

The overall mean amplitudes of the ASSRs in experiment 3 (Figure 7C) were 0.148/0.075/0.123/0.048 µV for 40/80/160/320 pps, respectively. The 160 pps showed significantly larger ASSRs compared to 80 and 320 pps. Significant correlations were also observed between the ASSR amplitudes of 160 pps and 80 pps, 160 pps and 320 pps, and 320 pps and 80 pps. In general, the correlations between N1P2 amplitudes and ASSRs were weaker in this experiment than in experiments 1 and 2. Only the offset N1P2 amplitudes and ASSRs for the 40 pps showed a significant correlation. No significant correlation was found between the upper pulse rate limit ($f_{\text{uplim}\_pps}$) and the N1P2 amplitude or the ASSR amplitude.

*Comparison of 40 Hz modulation rate conditions*

The ASSR for modulation frequencies up to 50 Hz is most likely generated from the auditory cortex (Herdman et al., 2002; Mäkelä and Hari, 1987; Pantev et al., 1994). The aim of this subsection is to compare the CAEPs or ASSRs evoked by the 40 Hz modulated SAM tones and the 40-pps filtered clicks.

▪ **CAEPs**

In experiment 3, we did not measure ACC2 data (inwards changes) for the filtered clicks, so only the onset, ACC1 (outwards changes), and offset responses of the five types of 40 Hz modulated SAM tones and the 40-pps filtered clicks were analyzed using GLMrm (with factors: stimuli type [400/800/1200/1600/4000SAM/40-pps-clicks], and response type [onset, ACC1 and offset]).

GLMrm showed a significant effect (p<0.005) of stimulus type, response type, and their interaction on the N1P2 amplitude. The mean amplitude was 7.544/6.271/5.860/4.759/5.046/2.309 μV for 400/800/1200/1600/4000SAM/40-pps-clicks, respectively. Pairwise comparison revealed significant differences between 400 Hz and 1600 Hz SAM tones, and between 40-pps filtered clicks and all four low-frequency SAM tones. Within each response type, the pairwise comparison showed that: 1) the onset response amplitude of the 40-pps filtered clicks was significantly smaller than most SAM tones except for the 800 Hz; 2) the 400 and 800 Hz SAM tones evoked significant larger responses than the other three SAM tones and the 40-pps filtered clicks for ACC1 response; 3) there were no significant differences among different stimulus types for the offset responses.

The GLMrm analysis revealed a significant effect of stimulus type, response type, as well as their interaction on N1 latency. The mean latency was 110/112/114/120/104/141ms for 400/800/1200/1600/4000SAM/40-pps-clicks. Pairwise comparison showed significant differences in latency between 1600 Hz and 4000 Hz SAM tones, and between 40-pps filtered clicks and the four low-frequency SAM tones. Further analysis within each response type showed that for onset responses, the 40-pps filtered clicks had a significantly different latency from most other stimulus types, except for the 1600 Hz SAM tones (p = 0.051). Within each stimulus type, there were no significant differences in latency among the three response types for both 4000 Hz SAM tones and 40-pps filtered clicks

▪ **ASSR**

To compare the 40-Hz ASSRs evoked by the 40-Hz modulated SAM tones and 40-pps filtered clicks within different analysis windows (see Supplementary Figure 3), GLMrm (with factors: stimuli type

[400/800/1200/1600/4000SAM/40-pps-clicks], and analysis window [T1, T2, and T4]) showed a significant effect of stimuli type, analysis window (p<0.001), and their interaction. The mean amplitude was for 400, 800, 1200, 1600, 4000 Hz SAM tones, and 40-pps filtered clicks were 0.168, 0.461, 0.142, 0.136, 0.091, and 0.154 $\mu$V, respectively. Pairwise comparison revealed significantly smaller amplitude of the 40-Hz ASSR evoked by the 4000 Hz SAM tones compared to the four low-frequency SAM tones, but no significant differences between the other stimulus types, including 4000 Hz SAM tones versus 40-pps filtered clicks. No significant differences were found between analysis windows T1 and T2, but as expected, T4 (silence) was significantly smaller than T1 and T2.

*Time-frequency domain*

As shown in Figure 4 and Figure 5, it is more challenging to determine the presence of detectable ACC responses evoked by the envelope ITD changes in high-frequency SAM tones and filtered clicks compared to those elicited by fine structure ITD changes in low-carrier frequencies. This requires more experience and possibly additional references, such as the overlying method demonstrated in Figure 5. To gain a better understanding of the smaller ACC responses evoked by envelope ITD changes and to enhance data visualization, Figure 8 displays the responses (with a high cutoff frequency of 1000 Hz instead of 30 Hz) in the time-frequency domain: SAM ITDfs (A), SAM ITDenv (B), filtered clicks IPTD (C). The average response from Figure 4 is re-plotted in each subplot as the black solid line around ~10 Hz. In general, the time-frequency amplitudes are related to stimulus parameters, such as $f_m$, and $f_c$.

The onset and offset responses in the time-frequency domain (Figure 8) displayed clear clusters of higher energy in the lower frequency range (<30 Hz) whenever there were detectable responses. In general, all three stimuli types evoked noticeable onset and offset CAEPs, except for the 40 pps filtered clicks.

The ACC responses evoked by ITD$_{FS}$ changes were easily recognizable in both response waveforms (Figure 4A) and the time-frequency domain (Figure 8A) for the 400, 800, and 1200 Hz SAM. However, it was smaller in 1200 Hz SAM and not distinguishable in the 1600 Hz in both the time and time-frequency domains.

In general, the ACC responses elicited by the envelope ITD changes were smaller compared to those evoked by the fine structure ITD changes. It appears there is greater presence of low-frequency brain activity (<10 Hz) for both types of high-carrier frequency stimuli in some cases. The ACC responses are roughly similar in scale to neighboring brain activities, making it challenging to distinguish an ACC response even in the time-frequency domain (Figure 8B and C).

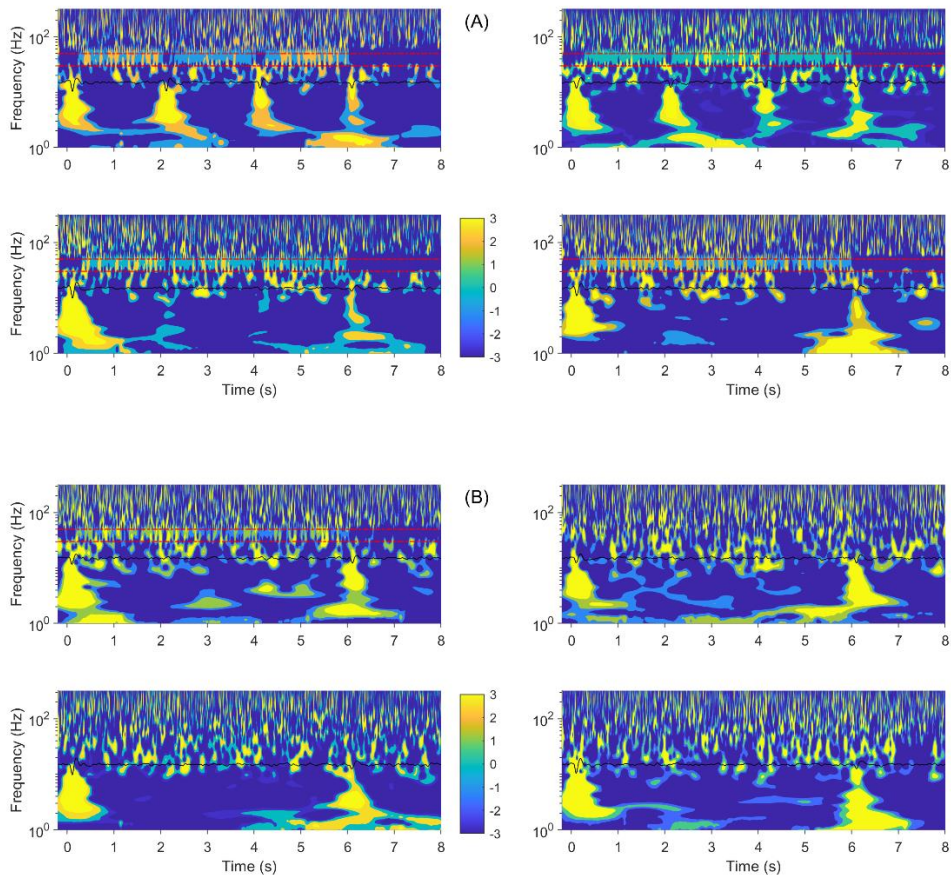

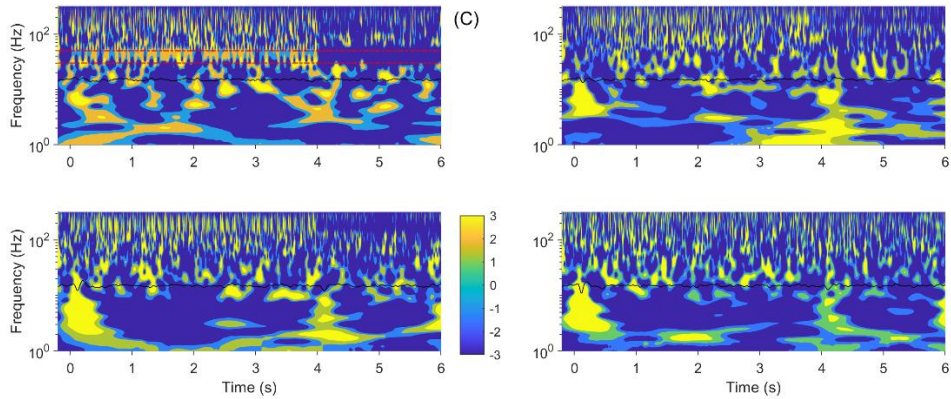

Figure 8 The average response in the time-frequency domain. The time-frequency spectrum was obtained through wavelet analysis of the average response, the black solid line shown in the center of each panel. The four conditions are shown in the upper left, upper right; bottom left, and bottom right panels: (A) $f_c$ = [200, 400, 800, 1600] Hz;. (B) $f_m$ = [40, 80, 160, 320] Hz. (C) pulse rate = [40, 80, 160, 320] pps. The two parallel red dashed lines correspond to 30 and 50 Hz.

Figure 8 reveals that the onset, ACC1, ACC2, and offset CAEPs had frequencies predominantly below 30 Hz, while the 40-Hz ASSRs (within the two parallel red dashed lines) were around 40 Hz except during the 2s-silence period. Time-frequency visualization shows possible interactions between the transient CAEPs and the ASSRs. Whenever the P1-N1-P2 complex was detectable and prominent, there was a noticeable reset (represented by the blue gaps in the 40 Hz regions) in the ASSRs. This suggests the steady-state activity was desynchronized by the prominent transient CAEPs, which might similar to the findings reported as 'interrupt' responses by some other groups (e.g., Bianco et al., 2020). For example, for $f_c$ of 400, 800, and 1200 Hz, the ASSRs were suppressed or reset around 0, 2, and 4s, respectively. There were notable energy differences in the ASSRs in these time windows (T1, T2, T3), as shown in Supplementary Figure 2 within the frequency range of 2-50 Hz.

## Discussions and conclusion

It has been reported in the literature that fine structure IPD- or ITD-based discrimination could be an important indicator for the performance in sound localization and the perception of speech related to phase locking in the auditory pathway (Eddins et al., 2018; Füllgrabe et al., 2015; Koerner et al., 2020; Moore, 2021; Papesh et al., 2017; Ross et al., 2007a; Ross et al., 2007b). In this study, CAEPs to IPD changes using an ACC paradigm was selected as a diagnostic assessment of binaural temporal processing. This paradigm has been demonstrated as robust and easily detectable in populations with various hearing profiles, both monaurally (Calcus et al., 2022; Han and Dimitrijevic, 2020; Mathew et al., 2017; Mathew et al., 2018; McGuire et al., 2021; Undurraga et al., 2020) and binaurally (Papesh et al., 2017; Ross et al., 2007a). Here, the ACC paradigm was simplified and recorded using a clinic-friendly EEG setup that can be conducted in less than one hour. Multiple responses were recorded simultaneously within the same EEG paradigm and analyzed. The same paradigm was used to compare the ITD sensitivity between low-frequency stimuli with fine structure ITDs and high-frequency stimuli with envelope ITDs. The EEG responses at different carrier frequencies and modulation rates were collected from the same participants within the same session. The response morphology of NH young participants was characterized as benchmark data for future studies involving the aging population and bilateral CI users.

In experiment 1 (fine structure ITD), 40 Hz modulated SAM tones with carrier frequencies of 1600 Hz or less were employed. The ACC evoked by IPD changes and the CAEPs evoked by the onset and offset of the SAM tones were recorded during the same session. Overall, the onset CAEPs and the ACC displayed similar morphologies. However, the mean peak latencies of ACC were generally longer than those of the onset and offset CAEPs (P1, 42/46/57/27 ms; N1, 114/132/137/95 ms; P2, 211/227/240/213 ms for onset/ACC1/ACC2/offset). Since the IPD change was introduced at the trough of the stimuli, the onset cues in the ACC response were minimized. Consequently, the present data suggested that the ACC responses were evoked by acoustic changes in the ongoing stimuli (e.g., IPD changes in experiment 1) rather than the onset of new stimuli. This supports the notion that the ACC is more than a simple onset response (Ostroff et al., 1998). However, the ACC exhibits differences and similarities with the onset and offset CAEP, indicating that these three responses may involve different but overlapping neural mechanisms.

In addition to the aforementioned three types of CAEPs, the ASSRs in the frequency domain were derived from the same data. The amplitude of the 40-Hz ASSRs gradually increased with increasing carrier frequency, consistent with Ross (2018). The mean ASSR amplitude was 0.187/0.173/0.152/0.142 for 400/800/1200/1600 Hz, respectively. The wavelet-based time-frequency visualization showed the interaction between detectable transient CAEPs and the steady state responses, particularly for the lower carrier frequency SAM tones (e.g., 400, 800, and 1200 Hz).

In summary, the method described in experiment 1 has potential as a tool for objectively evaluating the processing of changes in binaural information, especially regarding binaural temporal processing abilities. However, future research is needed to assess its variability (test-retest reliability) and relationship to behavioral tasks. The current study found no significant correlation between the EEG results and behavioral tasks (upper limit carrier frequencies). The lack of correlation could be due to difference in the procedures used in the psychoacoustic and EEG experiments. The behavioral tasks used an adaptive procedure to determine the upper-frequency limits, while most EEG data was not collected at the carrier frequency of the upper limit. Despite this, the results are encouraging and represent a step towards an objective measure that could be used to study binaural cue processing in a large population within a half-hour session. In clinical applications, the test conditions could be further reduced to shorten the required measurement time. The 2s stimuli was recommended for optimal clinical application of the ACC paradigm to provide a more robust response, which is consistent with the 1-2s longer stimuli recommended by Mathew (Mathew, 2018) and Calcus et al (2022). There is a time trade-off because fewer epochs would be possible in the same recording time.

A sample rate of 20 kHz was used in the EEG recording, which is much higher than the typical 1-2 kHz used in classic cortical EEG paradigms. In the post-processing, the EEG data was downsampled to various rates between 1-20 kHz to assess the minimum required sample rate for detecting transient CAEPs and ASSR. Although the average data was not significantly impacted by downsampling, the individual data showed differences between 1 kHz and higher rates (e.g. 8 and 20 kHz). Based on the results, a 4-channel EEG setup with a sampling rate above 1 kHz was recommended for clinical use in acoustic hearing assessments. If time and storage limitations are not an issue, higher channel numbers (e.g. 32 or 64 channels) and higher sampling rates (> 2 kHz) could offer the opportunity to employ more advanced post-processing techniques. However, further optimization may be required for electrical hearing assessments in CI users.

In experiments 2 and 3, high carrier frequency SAM tones and bandpass filtered clicks were used, respectively, to study the potential differences in ACC morphologies evoked by changes in envelope ITD compared to those evoked by fine structure ITD in normal-hearing participants. The use of bandpass filtered clicks was intended to mimic CI pulsatile stimulation and establish a foundation for future studies on CI users. The center frequency of 4000 Hz was chosen to eliminate or reduce phase locking in the fine structure, as envelope ITD changes with low carrier frequencies (< 2000 Hz) can primarily be attributed to fine structure changes.

Most participants in Ross (2018) showed no ACC responses to envelope ITD changes for the 4000 Hz $f_c$ SAM tones. Our findings confirmed Ross' results that the majority of ACC responses were at the same level as the noise floor, making it challenging to detect ACC responses. However, using wavelet time-frequency visualization (e.g., Figure 8) and overlaid plots of different test conditions (e.g., Figure 5), we are confident in the presence of ACC responses for modulation frequencies of 40, 80, and 160 Hz. Similar findings (small ACC responses) were obtained even for the bandpass filtered clicks. This suggests that the application of this paradigm to CI users is feasible, though it may pose difficulties, particularly due to the presence of CI stimulation artifacts (Hu and Dietz, 2015; Hu and Ewert, 2021; Hu et al., 2015). However, CI users may exhibit larger ACC responses than NH participants due to direct electrical stimulation of the auditory nerve. For example, clear electrically evoked brainstem binaural interaction component (ABR-BIC) has been recorded in bilaterally implanted animals (e.g., in cat, Hancock et al., 2010; Smith and Delgutte, 2007) and human CI users (Gordon et al., 2012; He et al., 2010; Hu and Dietz, 2015; Hu et al., 2016; Pelizzone et al., 1990), whereas acoustic ABR-BIC results are mixed. Some authors have reported an inability to measure acoustic BIC (Haywood et al., 2015), while others suggested it as a potential tool for binaural hearing tests (e.g., Benichoux et al., 2018; Brown et al., 2019; Riedel and Kollmeier, 2002; Ungan et al., 1997).

The ASSRs in experiments 2 and 3 were generally smaller than in experiment 1. The filtered clicks evoked larger ASSRs than high-frequency SAM tones, with the order of 40 Hz>160 Hz >80 Hz >320 Hz for both stimuli. The mean ASSR amplitudes for the 4000-Hz SAM tones modulated at 40/80/160/320 Hz were 0.093/0.058/0.071/0.022 µV, and for filtered clicks at 40/80/160/320 pps, they were 0.148/0.075/0.123/0.048 µV. The results suggest the ability to follow the high-frequency stimuli envelope at subcortical level, but the encoding of envelope information may differ between high and low frequency stimuli. The low-frequency fine structure ITD evoked higher ACC responses than the high-frequency envelope ITD (e.g., Figure 8), which also suggests potential differences in the central part. Recently, an article collected different viewpoints regarding the upper-frequency limit for the use of neural phase locking to code temporal fine structures in humans (Verschooten et al., 2019). Oxenham speculated that the human auditory cortex may process lower frequencies more effectively. A non-experienced participant reported finding the low-frequency SAM tones rhythmic, but the high-frequency stimuli "annoying" and "like background noise".

There is still ongoing debate about the origin of the two ITD sensitivity types, high frequency $ITD_{ENV}$ and low frequency $ITD_{FS}$. Some believe that the medial superior olives (MSO) neurons, which receive low-frequency excitatory inputs from both sides, are primarily sensitive to $ITD_{FS}$ (Goldberg and Brown, 1969; Yin and Chan, 1990), while ITD$_{ENV}$ sensitivity is believed to stem mainly from the excitatory-inhibitory interaction in the lateral superior olives (LSO) neurons, which are also sensitive to ILD (e.g., Dietz, 2016; Joris and Yin, 1995; Tollin, 2003; Tollin and Yin, 2005). This study aimed to obtain different CAEPs and ASSRs at different rates and bring laboratory findings close to possible clinical applications, rather than specifically addressing this question.

## Implications of possible applications and future work

The findings of this study have several important implications. Firstly, the used 4-channel clinical EEG setup with automatic P1, N1, and P2 peak-picking within the defined time window is promising and a sampling rate of at least 1 kHz is recommended for clinic use. Secondly, fine structure IPD changes can evoke ACC responses that can be recorded without participant involvement and are easily detectable with the 4-channel EEG setup. This has the potential as an objective tool for evaluating binaural sensitivity or for documenting binaural training effects. Thirdly, the small ACC responses evoked by the envelope ITD may pose a challenge when applying this paradigm to bilateral CI users, particularly in the presence of CI stimulation artifacts. However, these responses may be clearer in CI than in normal-hearing participants (e.g., Hu and Dietz 2015). Lastly, the researchers plan to test this paradigm in various populations with different hearing profiles to investigate differences in the neural encoding and processing of binaural and spatial cues. Further research is needed to characterize different types of CAEPs and ASSRs in CI users or participants with different degrees of neural deficits and to further develop the methods as a tool for remediating spatial processing deficits in these groups.

## CONFLICT OF INTEREST STATEMENT

The authors declare that the research was conducted in the absence of any commercial or financial relationships that could be construed as a potential conflict of interest.

## AUTHOR CONTRIBUTIONS

## FUNDING

This work was mainly funded by a Medical Research Council Senior Fellowship grant (MR/S002537/1) and partially funded by EMSATON (Projektnummer 415895050).

## ACKNOWLEDGEMENTS

The authors are grateful to Stephan Ewert for supporting the psychoacoustic experiment setup and Brain Moore for sharing their TFS-AF test and all the participants.

## DATA AVAILABILITY STATEMENT

Matlab implementation of the AFC framework is freely available at http://medi.uni-oldenburg.de/afc/. Other relevant data are presented within the paper

# Supplementary Materials

## Test procedures

Supplementary Figure 1 shows a flow chart of the performed experiments in this study.

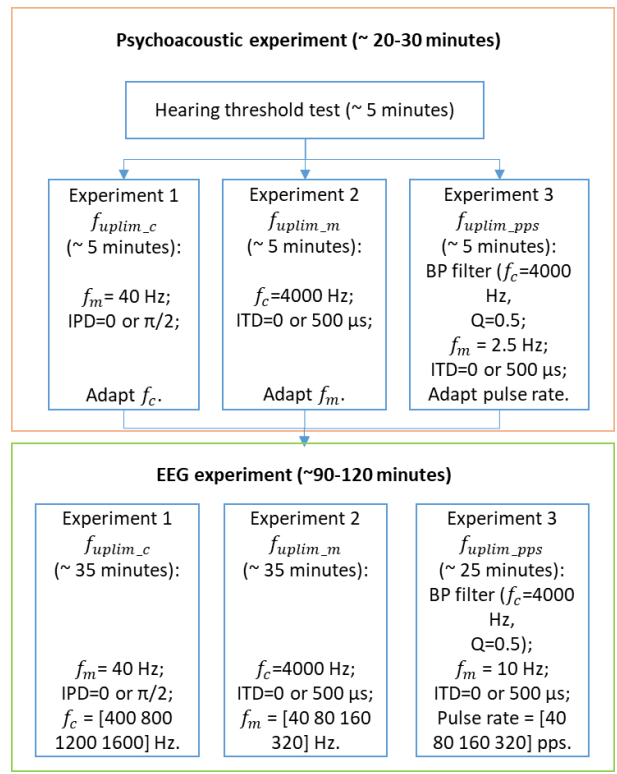

Supplementary Figure 1 (color online): The flow chart of the experiment plans. Participants first attend listening experiments (enclosed in the orange box, including an audiometry test, and three lateralization experiments) and then three EEG experiments (enclosed in the green box). The order of the three experiments in both psychoacoustic and EEG experiments was randomized for each participant.

## Time-frequency domain

Supplementary Figure 2A shows the time-frequency representations or wavelet spectrograms after wavelet transformations for stimuli with 40-Hz modulation frequency or 40-pps pulse rate in the frequency range of 2-50 Hz in the linear y-scale. The wavelet spectrograms reveal that the onset, ACC1, ACC2, and offset CAEPs are primarily dominated by frequencies below 30 Hz, while the 40-Hz ASSRs are centered around 40 Hz except during the 2s-silence period. Wavelet time-frequency visualization reveals some interactions between the transient CAEPs and ASSRs. A clear reset, represented by blue gaps in the 40 Hz regions, can be seen in the ASSRs whenever the P1-N1-P2 complex is detected and pronounced. This suggests that the transient CAEPs desynchronize the steady-state activity. For example, for $f_c$ of 400, 800, and 1200 Hz, the ASSRs were suppressed or reset at approximately 0, 2, and 4 s, respectively, and the ASSRs in these time windows (T1, T2, T3) exhibit notable energy differences.

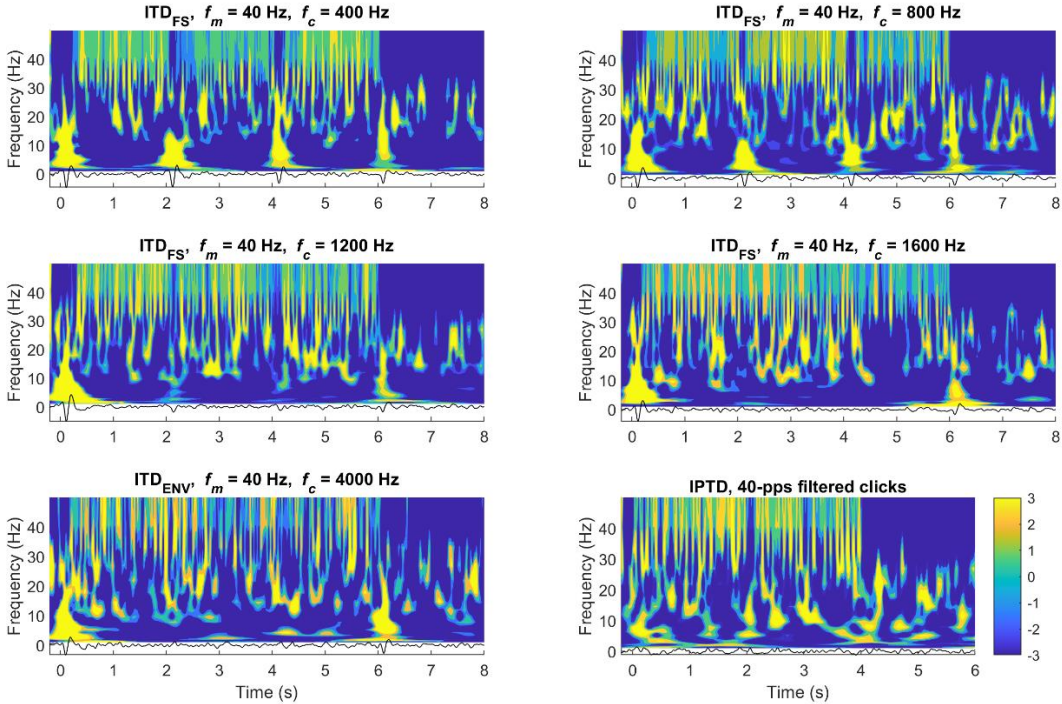

(A)

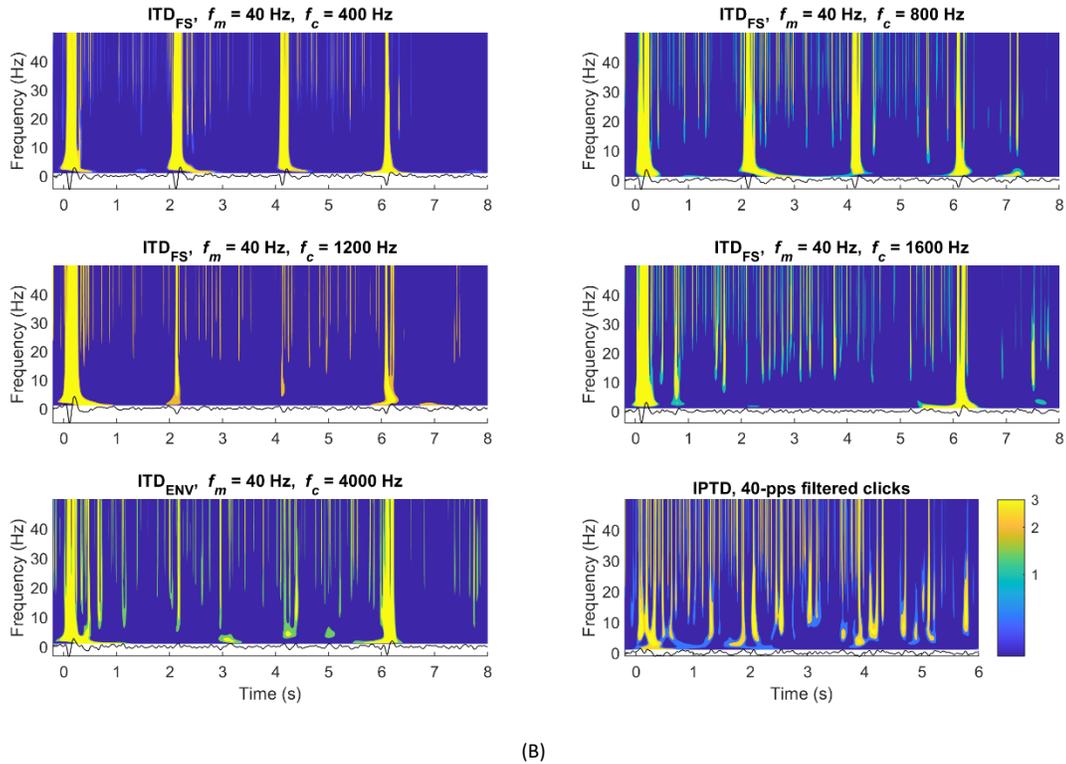

(B)

Supplementary Figure 2 The average response in the time-frequency domain for conditions with $f_m$ = 40 Hz or pulse rate = 40 pps. (A) the number of cycles is $n$ = 6, and (B) $n$ = 2, with log colorscale. The time-frequency spectrum was obtained by applying wavelet analysis on the average response shown at the bottom of each panel.

Both Figure 8 and Supplementary Figure 2A have n = 6 cycles, which determines the temporal and spectral precision. An increase in n leads to decreased temporal precision but increased spectral precision, and vice versa. To prioritize temporal precision, Supplementary Figure 2B presents the same results as Supplementary Figure 2A but with n = 2 cycles. The color bar uses a log color scale for improved visualization. Compared to the ACC evoked by low-frequency ITDfs, Supplementary Figure 2B shows that the change responses evoked by the high-frequency ITDenv are much smaller and more similar to the surrounding brain activities, making it more challenging to determine the presence of ACC response.

## Frequency domain (ASSRs)

A comparison between the 40-Hz ASSRs shown in Figure 8 (A, the top-left panel of B and C, with the two parallel red dashed lines representing 30 and 50 Hz, respectively) and Supplementary Figure 2 reveals slight differences in different time durations. To examine possible differences in the steady state responses before and after interruption by the stimulus onset, ITD changes, and offset, the ASSRs in different time windows were analyzed.

Supplementary Figure 3 shows the overall average ASSRs across participants for different analysis windows (panels 1-5: T1, T2, T3, T4, T1234, or T124). The colored curves in the cream area are the average ASSR from each individual. The red, blue, black, and pink curves represent the group average ASSR across participants for various test conditions: (A) $f_c$= [200, 400, 800, 1600] Hz; (B) $f_m$= [40, 80, 160, 320] Hz; (C) pulse rate = [40, 80, 160, 320] pps. The numbers with corresponding colors indicate the ASSR values at the modulation frequency within one of the analysis windows. The bottom right panel shows violin plots of the ASSR amplitude at the modulation frequency, within 8s (T1234) for the SAM tones or 6s (T124) for the filtered clicks.

In general, the 40-Hz ASSR decreased with increasing carrier frequency for the SAM tones. There was no ASSR in T4 (silence), the values shown are the noise floor around the modulation frequency, as expected.

Regarding experiment 1 (Supplementary Figure 3A), the overall mean 40-Hz ASSR amplitudes were 0.206/0.215/0.209/0.035/0.153 μV within T1/T2/T3/T4/T1234, and 0.187/0.173/0.152/0.142 μV for 400/800/1200/1600 Hz. As found by Ross (2018), the amplitude of the 40-Hz ASSR within T1234 declined gradually with increasing carrier frequency. GLMrm (factors: analysis window, T1, T2, T3, T4, T1234; $f_c$) showed significant effect of $f_c$ and analysis window, as well as their interaction. However, pairwise comparison showed no significant differences between different carrier frequencies. Pairwise comparision revealed significant differences between T4 (silence) and the other analysis windows (T1, T2, T3, T1234), as well as between T1234 and the other windows (T1, T2, T3, T4). Within each carrier frequency, the 40-Hz ASSR amplitudes were not significantly different among T1, T1, and T3.

The 40-Hz ASSR of different $f_c$ SAM tones were all correlated, but there was no correlation between the $f_c$ limit and either N1P2 or 40-Hz ASSR amplitude.

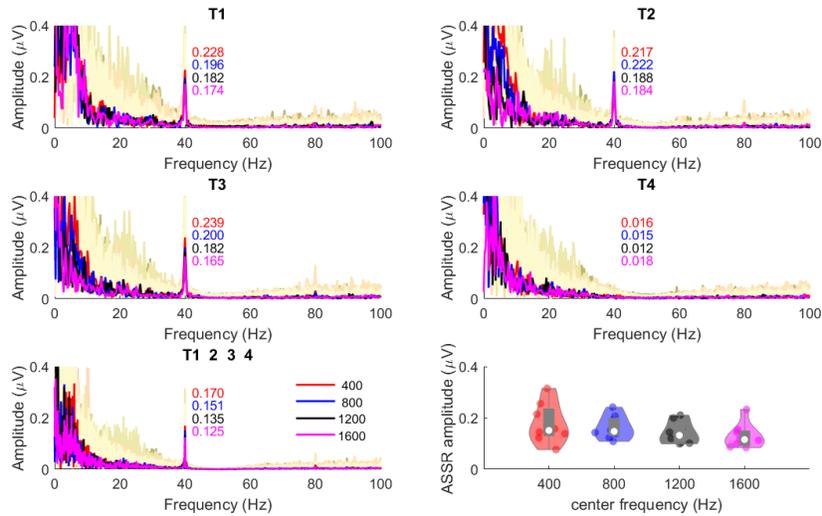

(A)

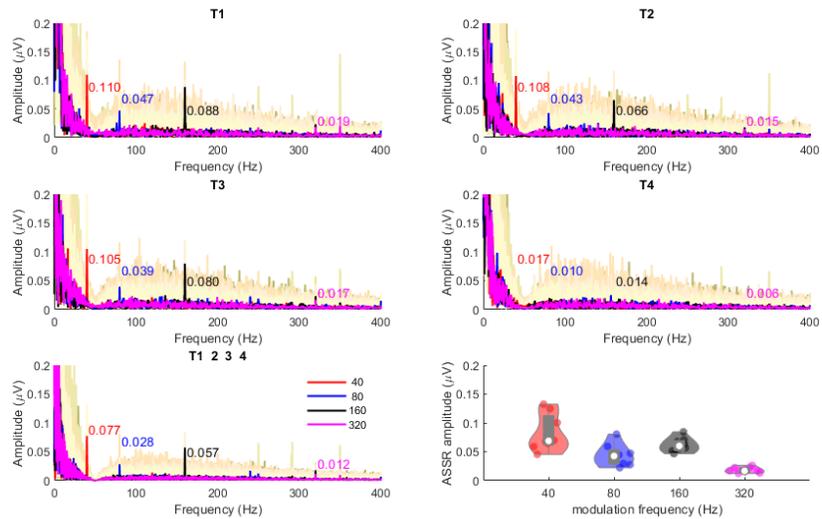

(B)

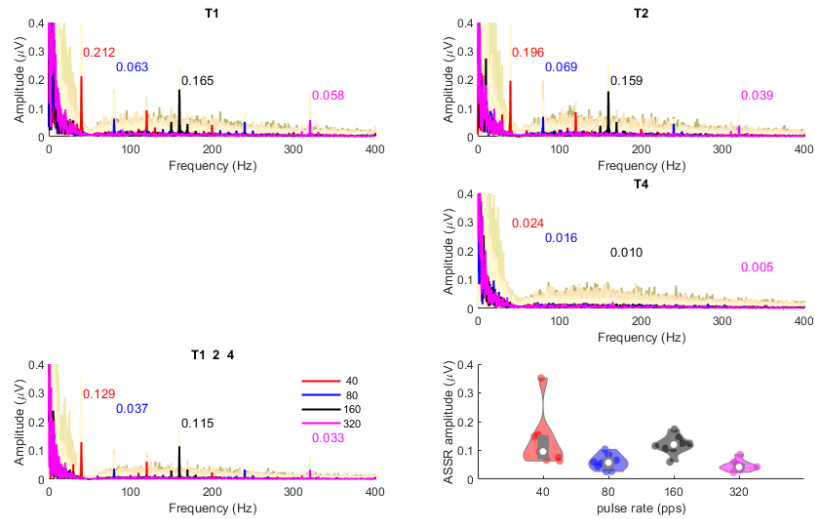

Supplementary Figure 3 The individual and group average ASSRs in the frequency domain. The red, blue, black, and pink spectrums are the overall average ASSR across participants for different test conditions: (A) $f_c$= [200, 400, 800, 1600] Hz;. (B) $f_m$= [40, 80, 160, 320] Hz. (C) pulse rate = [40, 80, 160, 320] pps. All the other colored curves in the background are the average ASSR from each individual. The first five panels are the ASSRs in different analysis time windows (T1, T2, T3, T4, T1234, or T124). The numbers shown in different colors are the 40-Hz ASSR values within the corresponding time window for each carrier frequency. The bottom right panel shows the violin plots of the ASSR amplitude for each carrier frequency, with an analysis window of 8s (T1234) for SAM tones or 6s (T124) for the filtered clicks. The solid dots in each violin plot are individual ASSRs at the corresponding $f_m$ or pulse rate of each participant.

Supplementary Figure 3B and C show the ASSRs at modulation rates of 40, 80, 160, and 320 Hz in various analysis time windows. The values in different colors indicate the corresponding ASSR values at different modulation rates. In general, the ASSRs of high carrier frequency stimuli are smaller than those of low carrier frequency (<=1600) SAM tones, and the ASSRs of filtered clicks are larger than those of high carrier frequency SAM tones. Among both types of high-frequency stimuli, the order is 40-Hz ASSR > 160-Hz ASSR > 80-Hz ASSR > 320-Hz ASSR. The larger 40-Hz ASSR evoked by the filtered clicks is mainly due to better phase locking to the envelope, as reported by Hu et al (2022).

Regarding experiment 2 (Supplementary Figure 3B), the overall mean ASSR amplitudes were 0.079/0.071/0.072/0.033/0.052 μV within T1/T2/T3/T4/T1234, and 0.093/0.058/0.071/0.022 μV for 40/80/160/320 Hz. GLMrm (factors: analysis window, T1, T2, T3, T4, T1234; $f_m$ ) showed significant effect of $f_m$ and analysis window, as well as their interaction. Pairwise comparisons showed no significant differences for $f_m$ 40 vs 80, 40 vs 160, 80 vs 160 Hz. Similar to experiment 1, pairwise comparisons showed significant differences between T4 (silence) and the other analysis windows (T1, T2, T3, T1234) as well as between T1234 and the other four analysis windows, both across modulation frequencies and within individual ones. Pairwise comparisons for each analysis widow showed that within T1, there were significant differences between most pulse rates except for 40 vs 160, 80 vs 160 pps; for T2, T3, T4, and T1234, the ASSR of 320 Hz was significantly smaller (T2, p<0.01; T3, T4, and T1234, p<0.05) than the other three modulation frequencies. There was no correlation between the $f_m$ limit and the N1P2 amplitude or the ASSR amplitude. The onset responses of different modulation frequencies were all correlated with each other, and the offset N1P2 amplitudes of 40, 80, and 160 Hz were correlated with each other, but not with that of 320 Hz.

For experiment 3 (Supplementary Figure 3C), the mean ASSR amplitudes were 0.143/0.131/0.032/0.089 μV in T1/T2/T4/T124, and 0.148/0.075/0.123/0.048 μV for 40/80/160/320 pps. GLMrm (factors: analysis window, T1, T2, T4, T124; $f_m$) showed significant effect of $f_m$ and analysis window, as well as their interaction. Pairwise comparison only showed significant ASSR amplitude differences for 160 vs 80, and 160 vs 320 pps. Unlike experiments 1 and 2, the pairwise comparison showed significant differences among all the analysis windows. Specifically, the 160 pps condition evoked a significantly larger ASSR compared to the 80 and 320 pps conditions within both T1 and T124, as well as the 320 pps condition within T2. Additionally, the noise floor in T4 (silence) was significantly smaller for 320 pps compared to the other three pulse rates. No correlation was found between the pulse rate upper limit and the N1P2 amplitude or the ASSR amplitude. However, the ASSR amplitude of the 40 pps condition was correlated with the 160 pps condition.